\pdfoutput=1
\documentclass[10pt,journal,compsoc]{IEEEtran}
\pdfoutput=1

\usepackage{amssymb,amsmath}
\usepackage{graphicx}
\usepackage{epstopdf}

\newcommand{\code}[1]{\texttt{#1}}
\usepackage{subfigure}
\usepackage{booktabs}
\usepackage{colortbl}
\usepackage{tabularx}
\usepackage{color}
\usepackage{xspace}
\usepackage{xcolor}
\usepackage{multirow}
\usepackage{float}
\usepackage{fancyhdr}
\usepackage[
  bookmarks=false,
  pdfpagelabels=false,
  hyperfootnotes=false,
  hyperindex=false,
  pageanchor=false
]{hyperref}
\hypersetup{
    colorlinks,
    linkcolor={black!50!black},
    citecolor={black!50!black},
    urlcolor={black!80!black}
}
\IEEEoverridecommandlockouts

\usepackage{tikz}
\usepackage{textcomp}
\usepackage{hyperref}
\usepackage{lipsum}

\newcommand\copyrighttext{%
  \footnotesize \textcopyright 2017 IEEE. Personal use of this material is permitted.
  Permission from IEEE must be obtained for all other uses, in any current or future
  media, including reprinting/republishing this material for advertising or promotional
  purposes, creating new collective works, for resale or redistribution to servers or
  lists, or reuse of any copyrighted component of this work in other works.  DOI:~10.1109/TMC.2017.2705680
  \lhead{Article Accepted in IEEE Transaction on Mobile Computing}
  \rhead{Early Draft}
  }
  
\newcommand\copyrightnotice{%
\begin{tikzpicture}[remember picture,overlay]
\node[anchor=south,yshift=10pt] at (current page.south) {\fbox{\parbox{\dimexpr\textwidth-\fboxsep-\fboxrule\relax}{\copyrighttext}}};
\end{tikzpicture}%
}
\begin{document}

\title{BRPL: Backpressure RPL for High-throughput and Mobile IoTs}
\author{Yad~Tahir,~
Shusen~Yang,~and
Julie~McCann
\IEEEcompsocitemizethanks{
\IEEEcompsocthanksitem Y. Tahir and J. McCann are with Department of Computing, Imperial College London.
\protect\\ E-mail: \{y.tahir11, j.mccann\}@imperial.ac.uk.
\IEEEcompsocthanksitem S.~Yang is with the  School of Mathematics and Statistics, Xi'an Jiaotong University, China. \protect\\
E-mail: shusenyang@mail.xjtu.edu.cn.
 }
\thanks{}}

\IEEEtitleabstractindextext{%
\begin{abstract}
RPL, an IPv6 routing protocol for Low power Lossy Networks (LLNs), is considered to be the de facto routing standard for the Internet of Things (IoT). However, more and more experimental results demonstrate that RPL performs poorly when it comes to throughput and adaptability to network dynamics. This significantly limits the application of RPL in many practical IoT scenarios, such as an LLN with high-speed sensor data streams and mobile sensing devices. To address this issue, we develop BRPL, an extension of RPL, providing a practical approach that allows users to smoothly combine any RPL Object Function (OF) with backpressure routing. BRPL uses two novel algorithms, \textit{QuickTheta} and \textit{QuickBeta}, to support time-varying data traffic loads and node mobility respectively. We implement BRPL on Contiki OS, an open-source operating system for the Internet of Things. We conduct an extensive evaluation using both real-world experiments based on the FIT IoT-LAB testbed and large-scale simulations using Cooja over 18 virtual servers on the Cloud. The evaluation results demonstrate that BRPL not only is fully backward compatible with RPL (i.e. devices running RPL and BRPL can work together seamlessly), but also significantly improves network throughput 
and adaptability to changes in network topologies and data traffic loads. The observed packet loss reduction  in mobile networks is, at a minimum, 60\% and up to 1000\% can be seen in extreme cases.

\end{abstract}

}

\maketitle
\copyrightnotice
\vspace{-2em}
\section{Introduction}

\IEEEPARstart{M}{anufacturers} are adapting a new breed of standards to provide an unprecedented level of transparency between business players, factory operations, and supply chain planning. This results in the fourth industrial revolution, commonly referred as ``Industry 4.0''. One of the core technological basis for this industrial revolution is the \textit{Internet of Things (IoT)}. Many efforts have been made to incorporate core Internet technology, TCP/IP standards, with emerging IoT standards primarily to ensure interoperability in heterogeneous networks. Herein nodes not only represent smart sensing devices, but also can be actuators, or even traditional Internet endpoints such as computers, tablets or smartphones.

The Internet Engineering Task Force (IETF) has proposed \emph{RPL} \cite{rd2009rpl} as a de-facto IoT routing standard for IPv6-based \emph{Low power and Lossy Networks (LLNs)}. RPL is a generic distance vector routing protocol that allows users to establish logical routing topologies, commonly known as \emph{Directed Acyclic Graphs (DAGs)}, over a shared physical network. DAGs are computed based on \emph{Objective Functions (OFs)} specified by users. Much work in both academia and industry has shown that RPL provides a promising routing solution for a wide range of network types and industrial applications, including  Home Automation (RFC 5826) \cite{brandt2010IETF}, Industrial Control (RFC 5673) \cite{pister2009industrial}, Urban Environments (RFC 5548) \cite{dohler2009routing}, Building Automation (RFC 5867) \cite{martocci2010building},  Advanced Metering Infrastructure (AMI) \cite{iyer2011performance}, and Smart Grids \cite{gungor2011}.

\subsection{Motivations}

To support real-world industrial IoT applications, the underlying networking services for LLNs must meet several requirements, including support for rapidly growing demands for \textit{high-throughput} networks \cite{quang2014throughput,moeller2010routing,chen2014big}, \textit{adaptability} to the data traffic dynamics~\cite{moeller2010routing,kafi2014congestion,sheng2013survey}, and in some cases \textit{mobility} of IoT devices~\cite{tunca2014distributedmobilesink,di2011datamobile,moeller2010routing}. Two of these requirements are formally stated in RFC 5673 \cite{pister2009industrial} and RFC 5867~\cite{martocci2010building}. However based on the current specifications, RPL can perform poorly in terms of meeting these requirements, which limits its adaption in numerous IoT applications.

\begin{figure}
 \centering
   \includegraphics[trim=140 370 420 80,clip=true,angle=0,width=0.30\textwidth] {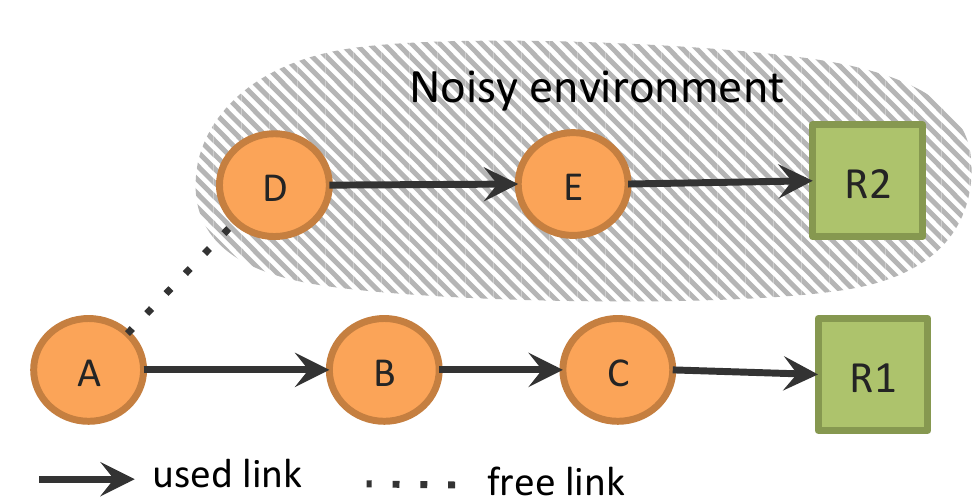}
  \caption{Used routing paths for a system with RPL and the ETX OF.}
  \label{fig:BRPLMotivation}
\end{figure}

1) \textbf{Throughput.} As a distance vector routing protocol \cite{medhi2010network}, RPL may suffer from severe congestion and packet loss when the network traffic is heavy (e.g., where multiple IoT applications coexist in a network). This is mainly because the DAGs defined by the OFs may not utilize the full network capacity. Fig. \ref{fig:BRPLMotivation} shows an example of this issue, in which node A wants to transfer data to the DAG's roots R1 or R2 (i.e., anycasting). Assume the network uses the standard ETX metric~\cite{de2005high} as the OF. In this case, RPL will use A-B-C-R1 in Fig. \ref{fig:BRPLMotivation} to avoid using the noisy path A-D-E-R2 as much as possible. As a result, congestion and packet loss may occur on path A-B-C-R1 when the network is handling a large amount of traffic, which is typical in multi-user IoT systems. This could be mitigated by exploiting the additional capacity of the suboptimal path A-D-E-R2.

2) \textbf{Traffic Load Adaptivity.} The highly dynamic and unplanned nature of LLN applications can produce time-varying data traffic patterns (e.g., traffic bursts generated by event-based applications). However, RPL fails to adapt to such traffic dynamics due to its fixed configurations. It is crucial to have an adaptive solution such that it utilizes necessary resources based on traffic demands. When the network experiences low data traffic, the default path specified by OF is sufficient. The suboptimal paths (e.g., path A-D-E-R2 in Fig.\ref{fig:BRPLMotivation}) are needed only when the data traffic volume is high.

3) \textbf{Mobility.} Due to the time-varying network topology caused by node mobility, invalid routes and link breakage are likely to exist in DAGs. The lack of adaptivity and mobility-awareness in an OF leads to the slow response to changes in network topologies. This makes RPL highly inefficient and potentially impractical in mobile networks.  Network resources must be utilized opportunistically as well as strategically to support low power and lossy IoT systems with mobile nodes.

\subsection{Contributions}

To address the mentioned limitations, Backpressure RPL (BRPL) is proposed, a new extension of RPL that provides enhanced support for high throughput, adaptivity and mobility without any modification or assumption on RPL OFs. Our contributions are summarized as follows:

1) We developed BRPL, the first work that incorporates the principles of backpressure-based optimization with RPL routing. BRPL is a multi-topology routing protocol that smartly routes traffic based on the gradients of both differential queue backlogs and OFs provided by the users. The basic idea of BRPL is to adaptively and smoothly switch between RPL and the backpressure routing (classic Lyapunov-based routing) according to network conditions. This is achieved by two lightweight control algorithms: \textit{QuickTheta} supporting  data traffic dynamics and \textit{QuickBeta} for topology dynamics (i.e., mobility).  BRPL has two interesting features: when the traffic loads in the network are  high, BRPL can achieve \textit{high throughput}  by utilizing all possible resources to handle the overwhelming data traffic. However, when the network experiences light data traffic in all nodes, BRPL is  \textit{objective function optimal} routing and becomes identical to RPL routing. As a result, BRPL maintains the advantages of both RPL and backpressure routing, while  avoiding their disadvantages.

2) BRPL uses the same control message structures defined in the specifications of RPL. This ensures that BRPL is \textit{interoperable} with RPL, i.e. IoT devices running original RPL or BRPL can operate together seamlessly in a hybrid network, without requiring any source code modification on nodes running RPL. Such software compatibility is very important in practice, as the routing layer of some nodes (such as non-programmable Zigbee chips) is not reprogrammable or replaceable. To the best of our knowledge, this is the first work that aims to make backpressure-based routing feasible in hybrid networks.

3) BRPL strongly adheres to the ``\textit{Application Transparency}'' design principle. Improving throughput, adaptivity and mobility of the network in BRPL is achieved without requiring any knowledge or statistical assumptions on OFs and their implementation details. This is vital in IoT as OFs for IoT applications can be widely dissimilar. The current specifications of RPL do not provide any constraint on the types of OFs.

4) We implemented BRPL in Contiki OS~\cite{contiki}, an open source operating system for LLNs and IoT. Through both real-word experiments on the FIT IoT-LAB testbed \cite{iotlab} and extensive simulations with Cooja (Contiki's network simulator) over a private Cloud with 18 servers, we demonstrate that BRPL can work seamlessly with RPL, and it has control overhead similar to RPL and backpressure routing. More importantly, the results show that BRPL smartly achieves the advantages of both RPL and backpressure routing and significantly outperforms them in terms of reliability, end-to-end delay, throughput, adaptability to network dynamics, and mobility support.

\subsection{Paper Organization}

The remainder of the paper is organized as follows. The next section presents system models and more detailed information about RPL. BRPL extension is proposed in Section \ref{sec:algo}. Section \ref{sec:analysis} provides a detailed discussion about our proposed solution. Testbed experiments and simulation results are presented in Section \ref{sec:evaluation}. Related work is discussed in Section \ref{sec:related_work}. Finally, we conclude the paper in Section \ref{sec:conclusion}.
\linebreak

Table \ref{tab:symbols} summarizes the key symbols used in this work.

\begin{table}[h!]
\centering
\caption{Main symbols used in this paper.}
\begin{tabular}{|l|l|}
\hline
$\mathcal{S},\mathcal{R}$ & The sets of all sensor nodes and roots, respectively.\\ \hline
$\mathcal{N}$ & The set of all IoT devices $\mathcal{N}=\mathcal{S}\cup\mathcal{R}$.\\ \hline
$\mathcal{L}$ & The set of all wireless links.\\ \hline
$\mathcal{N}_{x}(t)$ & The set of $x$'s all neighbors at slot $t$.\\ \hline
$\mathcal{M}$ & The set of all DAGs available in the network.\\ \hline
$\mathcal{R}$ & The set of all roots for the network. \\ \hline
$\mathcal{R}_m$ & The set of all roots for DAG $m$.\\ \hline
$c_{x,y}(t)$ & Channel capacity of link $x, y$ at slot $t$.\\ \hline
$f^m_{x,y}(t)$ & Data rate for DAG $m$ over link $(x,y)$ at  $t$.\\ \hline
$f_{x,m}^{in}(t)$ & Total incoming DAG $m$ data rate of node $x$ at $t$.\\ \hline
$f_{x,m}^{out}(t)$ & Total outgoing DAG $m$  data rate of node $x$ at $t$.\\ \hline
${RootRank}_x^{m}$ & The Rank of root $x$ for DAG $m$.\\ \hline
$Rank_{x}^{m}(t)$ & The Rank of node $x $ for DAG $m$  at $t$ .\\ \hline
$p_{x,y}^{m}(t)$ & The penalty over link for DAG $m$   $x, y$ at $t$.\\ \hline
$Q_{x}^{m}(t)$ & The queue length of node $x$ for DAG $m$ at $t$.\\ \hline
$r^m_x(t)$ & Data packets generated by node $x$ for DAG $m$ at $t$. \\ \hline
$Q_{y}^{m}(x,t)$ & The queue length for the neighbor record $y$\\
              & in the neighbor table of node $x$. \\ \hline
$MaxQ_x^m$ & Maximum queue size for DAG $m$ in node $x$.\\ \hline
$MaxQ_y^m(x)$ & Maximum queue size for the neighbor record $y$\\
              & in the neighbor table of node $x$. \\ \hline
$w_{x,y}^{m}(t)$ & The weight over link $(x, y)$ for DAG $m $ at $t$\\ \hline
$\theta_{x}^m(t)$ & Tradeoff parameter between throughput and OF.\\ \hline
$\triangle Q_{x,y}^{m}(t)$ & the queue length difference between $x, y $ for DAG $m $.\\ \hline
$\bar{Q}_{x}^{m}(t)$ & The moving average of $x$'s queue for DAG $m$ at $t$.\\ \hline
$\beta_x(t)$ & The mobility-awareness parameter for node $x$ at $t$.\\ \hline
\end{tabular}

\label{tab:symbols}
\end{table}

\newpage
\section{System Models}

\begin{figure}
 \centering
   \includegraphics[trim=45 15 272 15,clip=true,angle=0,width=0.45\textwidth] {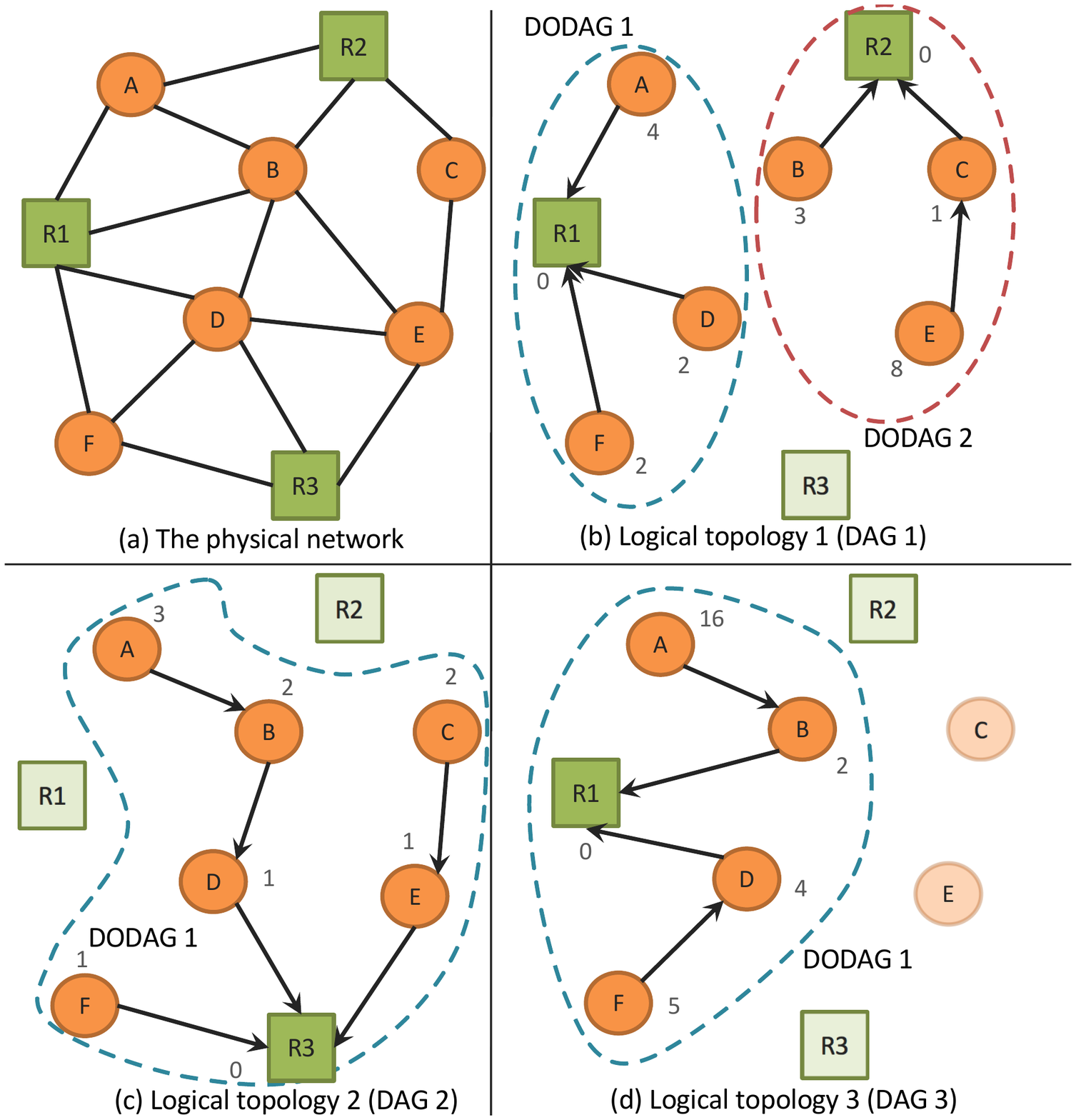}
  \caption{An illustration for logical routing topologies in RPL. The circles and squares represent nodes and roots, respectively. Ranks are represented as numbers next to the nodes and roots. Both (c) and (d) topologies have one DODAG and one DAG each, whereas the logical topology in (b) has two DODAGs and one DAG.}
  \label{fig:RPLillustration}
\end{figure}

We consider an LLN that consists of a set of IoT devices $\mathcal{N}=\mathcal{S}~\cup~\mathcal{R}$  communicating in a multi-hop fashion as shown in Fig. \ref{fig:RPLillustration}(a); where $\mathcal{S}$ represents the set of all sensor nodes that can generate and relay data packets, and $\mathcal{R}$ is the set of all roots/gateways that collect the data traffic produced by the network. The network operates in discrete time slots $t \in \{1,~2, ...\}$.

\subsection{Modeling the Physical LLN}

To model the dynamic and lossy wireless transmissions,  we define ${\rm c^{max}}\geq c_{x,y}(t) \geq 0$ as the logical link-layer channel capacity from node $x$ to node $y$ at time slot $t$, i.e. $c_{x,y}(t)$ indicates the maximum (integer) number of acknowledged packets that can be\textit{ successfully transmitted} from node $x$ to $y$ during time slot $t$.  Here, ${\rm c^{ max}}$ is the maximum possible $c_{x,y}(t),\forall t$, which is bounded by the data rate of the wireless radio. For instance, experimental studies show that a commonly-used IEEE 802.15.4 transceiver, CC2420 (e.g.\cite{micaz}), can achieve a data rate of approximate 160 40-bytes packets per second \cite{sridharan2009explicit} in practice.

The logical channel capacity depends on a wide range of random events such as wireless interference and node mobility. When $c_{x,y}(t) > 0$, it indicates that both node $x$ and $y$ are one-hop away from each other at time slot $t$. Let $\mathcal{N}_{x}(t) \subseteq \mathcal{N}$ be the set of all possible one-hop neighbors that node $x \in \mathcal{N}$ can communicate with during slot $t$:

\begin{equation*}
\mathcal{N}_{x}(t) := \{\:y\:|\:c_{x,y}(t) > 0,\:c_{y,x}(t) > 0,\:y \in \mathcal{N}\:-\:\{x\}\:\}
\end{equation*}

The whole network is modeled as a time-varying weighted graph $G(\mathcal{N},\mathcal{L},\mathbf{c}(t))$ where $\mathcal{L}$ represents all possible wireless links for all node pairs in $\mathcal{N}$.
 $|\mathcal{L}|$--dimensional vector $\mathbf{c}(t)$ holds the channel capacities for all the links at $t$.

\subsection{ Network-theoretical Modeling of RPL routing}

The IETF Routing over Lossy and Low-power Networks (RoLL) group established the specifications of RPL,  published as RFC 6550 \cite{rd2009rpl}. This section presents network theoretic models for the key RPL specifications.

\subsubsection{ Multi-Commodity  Model for Multi-Topology Routing}
Topologically, RPL adheres to the concept of \textit{Multi-Topology Routing (MTR}). RPL allows multiple instances of RPL to run concurrently in a network. Each RPL instance constructs a logical topology called \textit{Directed Acyclic Graph (DAG)} that  consists of one or more \textit{Destination Oriented DAGs (DODAGs)} (one DODAG per one root destination). All logical DAGs share the same physical network infrastructure, and simultaneously support different routing optimization objectives.
For instance, Fig.\ref{fig:RPLillustration}  shows illustrative examples of three DAGs for a common physical LLN, where the DAG in Fig.\ref{fig:RPLillustration} (b) consists of two DODAGs, and the DAGs in Fig.\ref{fig:RPLillustration} (c) and (d) have one DODAG each.

Let $\mathcal{M}$ be  the set of all DAGs available in the network. $\mathcal{R}_m$ represents the set of all roots for the DAG $m \in \mathcal{M}$. Hence, the set of all roots for all DAGs can be seen as:

\begin{equation}
\mathcal{R}=\bigcup_{m\in\mathcal{M}}\mathcal{R}_{m}
\end{equation}

We use the multi-commodity model \cite{hu1963multi} to formalize the RPL protocol in a network. Here, each DAG $m$ in the network can be seen as a commodity.  All data traffic in a DAG $m$ should be transmitted to any destination $r\in\mathcal{R}_m$.
\vspace{0.3em}
\\\textbf{Definition 1}~[MTR Traffic Feasibility]. \textit{Denote  $ f^m_{x,y}(t)\geq0$ as the actual amount of  data traffic for DAG $m$ over a wireless link $(x,y)$ at slot $t$. Let $f_{x,m}^{in}(t)=\sum_{y\in\mathcal{N}_x(t)}f^m_{y,x}(t)$ and $f_{x,m}^{out}(t)=\sum_{y\in\mathcal{N}_x(t)}f^m_{x,y}(t)$ be the total data DAG-$m$ incoming and outgoing traffic of node $x$ at slot $t$ respectively. Then for any feasible multi-topology routing approach (e.g. RPL),  the following two conditions should be satisfied:}
\begin{eqnarray}
\sum_{m\in\mathcal{M}}f^m_{x,y}(t)&\leq &c_{x,y}(t),\forall x,y\in \mathcal{N}, t\geq1\label{eq:capacityconstraint}\\
\overline{r}^m_{x}+\overline{f}_{x,m}^{in}&\leq&\overline{f}_{x,m}^{out},\forall x\in \mathcal{S}, m\in \mathcal{M}\label{eq:flowconstraint}
\end{eqnarray}
where  $\overline{f}_{x,m}^{in}$ and $\overline{f}_{x,m}^{out}$ represent the long-term averages of  ${f}_{x,m}^{in}$ and ${f}_{x,m}^{out}$ respectively, and  $\overline{r}^m_{x}\geq0$ is the long-term average of DAG-$m$ data traffic generated by node $x$.

 Condition (\ref{eq:capacityconstraint}) ensures that the total data traffic for all DAGs over a link should not exceed its channel capacity. Condition (\ref{eq:flowconstraint}) states the \textit{flow conservation law}, i.e. the total incoming data traffic for a given DAG $m$  at any IoT device $x$ should not be more than the total outgoing data traffic for DAG $m$ for $x$.

\subsubsection{Objective Function and Routing Gradient}
\begin{figure}
 \centering
   \includegraphics[trim=70 320 310 70,clip=true,angle=0,width=0.45\textwidth] {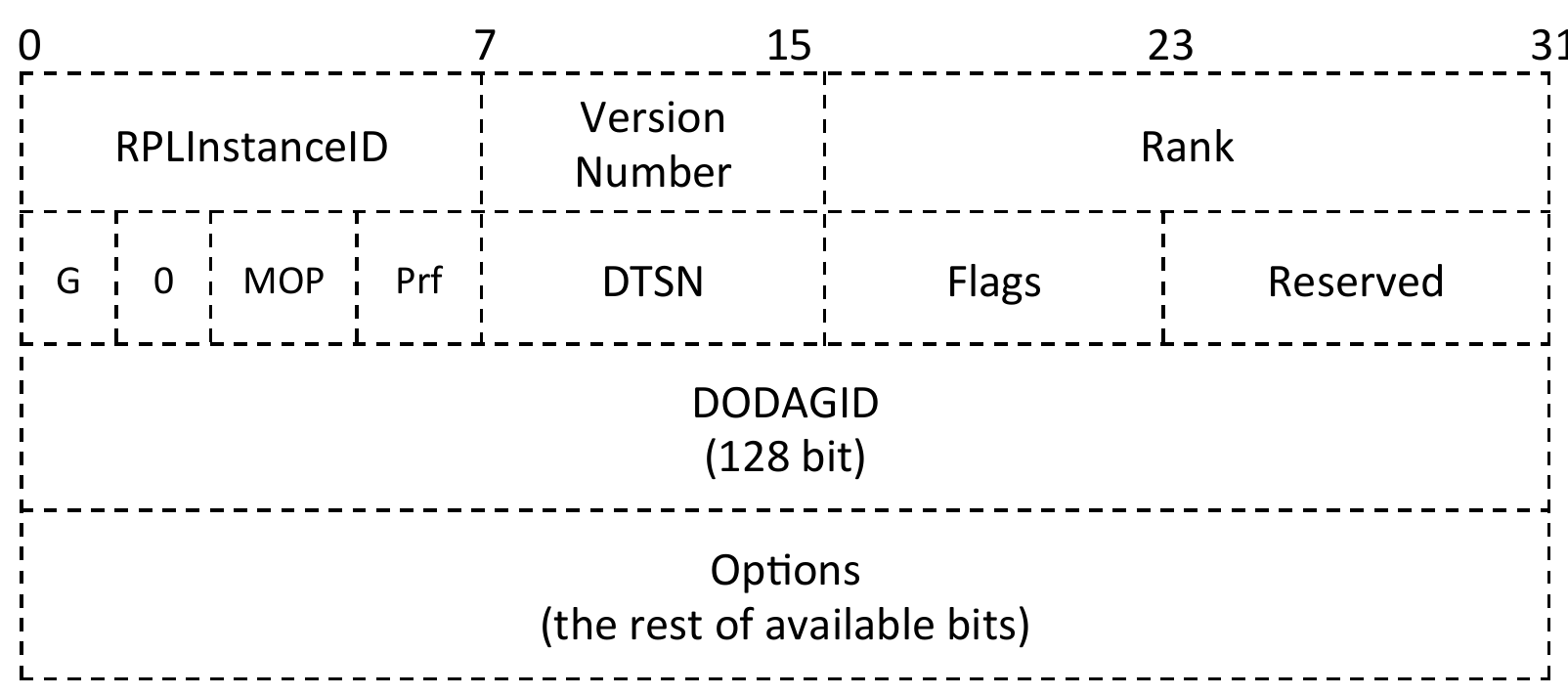} 
  \caption{DIO message  structure in RPL.}
  \label{fig:DIOStrcture}
\end{figure}

RPL constructs each DAG by using an OF, which defines a specific routing optimization objective, such as minimizing  energy consumption or end-to-end delay. All DODAGs within a DAG share the same OF.
Each  DODAG  in a DAG is computed by a scalar variable associated with each node called \code{Rank}, which is basically the logical distance between a node and the corresponding root of the DODAG. For instance, the Rank in Fig.\ref{fig:RPLillustration}(c) represents hop count, and the routing objective is to minimize the number of hops for packet delivery from the nodes to the destination R3.
The Rank value of node $x\in\mathcal{N}$ in $m\in\mathcal{M}$ is computed as:

\begin{equation}
Rank_{x}^{m}(t) = \left\{
   \begin{array}{l l}
    \underset {y \in \mathcal{N}_{x}(t)} {\min}( p_{x,y}^{m}(t) + Rank_{y}^{m}(t))& \quad x\notin\mathcal{R}_{m} \vspace{0.2em}\\
    {\rm RootRank}_x^{m} & \quad x\in\mathcal{R}_{m}
   \end{array} \right.
\label{eq:rplRank}
\end{equation}
where $p_{x,y}^{m}(t) > 0$ denotes the penalty/cost of using the link $(x,y)$ in DAG $m$ at time slot $t$, and ${\rm RootRank}_x^{m} \geq 0$ is the smallest Rank value in the DAG $m$. Hence, when a node $x$ is not a root, its rank is computed by finding a one-hop connection that gives the smallest sum of neighbor rank $Rank_{y}^{m}$ and link penalty $p_{x,y}^{m}$.

\subsubsection{DIO Message and DODAG Establishment}

The construction of any DODAG \footnote{For brevity, this work only considers RPL nodes operating as both leaf and router. However, it is straightforward to extend our proposed solution to support the leaf-only and router-only modes.} is fully distributed, but initially triggered by the root $r$ of the DODAG. $r$ starts by broadcasting a DODAG Information Object (DIO) message to its one-hop neighbors. Fig. \ref{fig:DIOStrcture} that shows the DIO message structure. DIO is an ICMPv6 information message holding important parameters about the DODAG including: RPLInstanceID, Version Number, and the Rank of the sender.

Each neighbor $x\in\mathcal{N}_{r}(t)$ computes its Rank value based on Eq.(\ref{eq:rplRank}),  updates the Rank parameter, and then broadcasts its DIO message to its one-hop neighbors $y \in \mathcal{N}_{x}(t'), t' \geq t$. This process repeats itself for all other nodes existing in the network. Based on the computed Rank values, each node $x$ is going to choose its optimal neighbor $y^*_m$ {as follows}:

\begin{equation}
y^*_m(x,t)=\arg\min_{y\in\mathcal{N}_x(t)} ( p_{x,y}^{m}(t) + Rank_{y}^{m}(t))
\label{eq:bestneighbor}
\end{equation}

This neighbor is commonly referred as the \textit{preferred parent} for node $x$ in the DAG $m$. Whenever a node receives data packets associated with $m$ DAG, it forwards them to its preferred parent. The parent then repeats the process until a root $r\in\mathcal{R}_m$ receives the packets.

The broadcasting process for RPL relies on the Trickle algorithm specified in RFC 6206. It is not necessary to have a broadcasting process per each DODAG, instead it can be per each DAG. It is important to note that the Trickle algorithm takes network stability into account. Receiving a DIO message from a sender with a lesser Rank that causes no changes to the recipient's preferred parent or Rank is considered as a `consistent' DIO message with respect to the Trickle timer. When the network continuously encounters consistent DIO messages (i.e. the network is stable), the algorithm decreases the broadcasting rate, which results in performing less update operations on the neighbor tables. In this case, the tradeoff between energy efficiency and neighbor table consistency is considered to be highly justifiable for LLNs, particularly when nodes have limited resources. However, in time-varying networks, the Trickle algorithm increases the DIO broadcasting rate. Neighbor table maintenance is therefore performed more frequently to ensure DAG consistency and accuracy.

\section{BRPL}
\label{sec:algo}

\begin{figure}
 \centering
   \includegraphics[trim=50 225 50 80,clip=true,angle=270,width=0.37\textwidth] {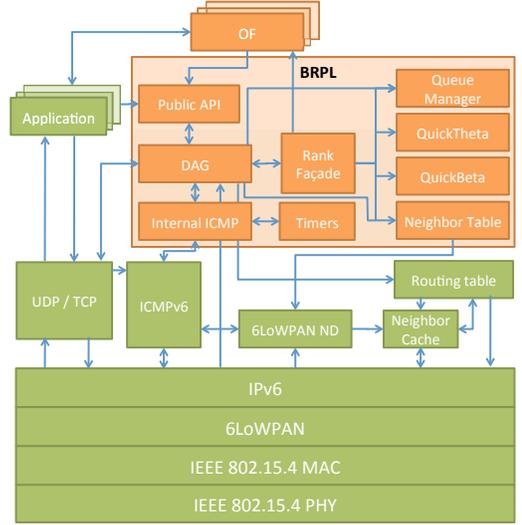} 
  \caption{Illustration of software architecture of BRPL.}
    \label{fig:BRPL_ARCH}
\end{figure}

This section describes our proposed extension to RPL, BRPL, aiming to enhance the performance of RPL in terms of mobility support, high throughput, and adaptivity to the network traffic dynamics. The software architecture of the proposed solution, which is illustrated in Fig. \ref{fig:BRPL_ARCH}, has the following key components:

\begin{itemize}
 \item\code{Internal ICMP} communicates with ICMPv6 in order to send/receive DIO, DAO, and DIS control messages. When an ICMPv6 message is received, the internal ICMP component first extracts the payload, and then notifies the DAG component.
  \item\code{Timers} holds all timer-related logics including the Trickle algorithm.
  \item\code{Public API} provides a clean interface to allow external components such as user applications to adjust or interact with BRPL.
\item \code{QuickBeta} holds the implementation details for the mobility-awareness indicator.
\item \code{QuickTheta} an online algorithm that actively adjusts the parameter settings of BRPL based on current network dynamics.
 \item\code{OF} is the  objective function provided by the application layer.
 \item\code{Neighbor Manager} manages the neighbor table of the node. It communicates with IPv6 Neighbor Discovery Service to synchronize the neighbor table.
 \item\code{Queue Manager} manages the data buffer (queue) of each DAG. The queue stores incoming IPv6 data packets.
 \item \code{Rank Fa\c{c}ade} is responsible for calculating Ranks and link weights for the one-hope neighbors. As the name indicates, this component follows the well-known ‘Fa\c{c}ade’ design pattern \cite{gamma1994design} in software engineering.
  \item\code{DAG} contains all the logic related to DAGs including modifying IPv6 routing table, routing repair, choosing best parent based on \code{Rank Fa\c{c}ade}.
\end{itemize}

\subsection{Multi-topology Queueing System}
Similar to RPL, BRPL follows the design principles of MTR. Each DAG is established by an OF.  However unlike RPL, BRPL combines network congestion gradients (i.e. differential queue gradients) and OF ranking for handling upward data routing. Each IoT device running BRPL is required to maintain a queue (packet buffer) for each DAG (This queue is maintained by the queue manager in Fig. \ref{fig:BRPL_ARCH}). Let $Q_{x}^{m}(t) \geq 0$ be the queue backlog (or queue length) of node  $x\in\mathcal{N}$  for  DAG $m\in \mathcal{M}$ at slot $t$. The queue dynamics of $x$ are defined as follows:
\begin{eqnarray}
 &&0\leq Q_{x}^{m}(t) \leq MaxQ_{x}^{m}\\
 &&Q_{x}^{m}(t+1)=|Q_{x}^{m}(t)-f_{x,m}^{out}(t)|_{+}+r^m_x(t)+f_{x,m}^{in}(t)\qquad
\end{eqnarray}

where $MaxQ_{x}^{m}>0$ is the maximum queue length (i.e. allocated data buffer size) of $x$ for DAG $m$.  $r^m_x(t)$ denotes  the amount of data packets produced by node $x$ for DAG $m$ at time slot $t$. The operator $|a|_{+}$ means $\max(a,0)$. The queue length of any root is always equal to zero:

\begin{equation*}
 Q_{r}^{m}(t)=0,~\forall r\in\mathcal{R},~m\in \mathcal{M},~t\geq1
\end{equation*}

\subsection{Neighbor Table Maintenance}
When a node receives a DIO message from a one-hop neighbor, it updates a few fields for the sender's record in the neighbor table. This includes \textit{rank}, \textit{queue backlog} and \textit{maximum queue length}.

Because this work considers hybrid networks (i.e. some nodes may use the original specifications of RPL and do not advertise queue backlogs), the `queue backlog' field for any neighbor record in BRPL is updated as follows:

\begin{equation}
Q^m_y(x,t)=\\
\left\{
   \begin{array}{l l}
    Q^m_y(t) & ~y ~{\rm is~a~BRPL~node} \\\\
    \frac{Rank^m_y(t)}{Rank^m_x(t)}Q^m_x(t) &~y~ {\rm is~a~RPL~node}
   \end{array} \right.
   \label{eq:updateQueueNeighborTable}
\end{equation}

Where $Q^m_y(x,t)$ denotes the queue backlog (a counter) for the node $y$ in $x$'s neighbor table. Eq. \ref{eq:updateQueueNeighborTable} is designed carefully to address hybrid networks. Here we have two cases:

When $y$ runs the original RPL routing protocol, then queue  details are going to be missing in $y$'s DIO messages. In this case, node $x$ updates the $Q^m_y(x,t)$ by scaling its queue length $Q^m_x(t)$ based on the rank of $x$ and $y$. The use of the ${Rank^m_y(t)}/{Rank^m_x(t)}$ ratio avoids $x$ sending data packets to its child nodes, thus less routing loops. Based on Eq. \ref{eq:rplRank}, we can observe that nodes choosing $x$ as a parent have ranks equal to or larger than node $x$'s rank. 

The second case is when $y$ runs BRPL. Here, $Q^m_y(t)$ is not missing in DIO messages. $x$ updates the $Q^m_y(x,t)$ field such that $Q^m_y(x,t)=Q^m_y(t)$.

Likewise, the maximum queue length field $MaxQ_{y}^{m}(x)$ for neighbor $y$ in $x$'s neighbor table is updated according to:

\begin{equation}
MaxQ_{y}^{m}(x) =\\
\left\{
   \begin{array}{l l}
    MaxQ_{y}^{m} & ~y ~{\rm is~a~BRPL~node} \\
    MaxQ_{x}^{m} &~y~ {\rm is~a~RPL~node}
   \end{array} \right.
\end{equation}

BRPL also uses the Trickle algorithm to control the broadcasting rate of DIO messages. Similar to RPL, BRPL is a fully distributed routing protocol. Each node only needs to broadcast its DIO messages to its one-hop neighbors. There is no need to relay DIO messages for other nodes. Global repairing operations are generally not required.

\subsection{Data Forwarding based on RPL Ranks and Queue Backlogs}
At each time slot $t$, BRPL performs routing and data forwarding operations for upward data traffic as follows:

\begin{itemize}
\item\textbf{Link Weight Calculation.} Each IoT device $x$ running BRPL computes a weight $w_{x,y}^{m}(t)$ for each neighbor $y\in\mathcal{N}_x(t)$ by combining queue length information and RPL rank values:

\begin{equation}
w_{x,y}^{m}(t)=  \theta_{x}^m(t)\tilde{p}_{x,y}^{m}(t)-(1-\theta_{x}^m(t)) \triangle Q_{x,y}^{m}(t)c_{x,y}(t)
\label{eq:RankMRPL}
\end{equation}
where
\begin{eqnarray}
&& \tilde{p}_{x,y}^{m}(t)= p_{x,y}^{m}(t)+Rank^m_y(t)\\
&&\triangle Q_{x,y}^{m}(t) = Q^m_x(t) - Q^m_y(x,t)
\end{eqnarray}
and $0 \leq \theta_{x}^m(t) \leq 1$ is the tradeoff parameter between average queue backlogs and minimizing objective function of DAG $m$, which is adaptively updated in real-time by the QuickTheta algorithm\footnote{Please note that the weight $w_{x,y}^m(t)$ for BRPL adopts the typically combination of queue backlog difference $\triangle Q_{x,y}^{m}(t)$ and penalty $\tilde{p}_{x,y}^{m}(t)$, which are also used in other backpressure based routing algorithms such as \cite{moeller2010routing,yang2013selfish}.}.

However in practice, the Eq. (\ref{eq:RankMRPL}) may suffer from \textit{scaling} issues between the range values of $\triangle Q_{x,y}^{m}(t)$ and $\tilde{p}_{x,y}^{m}(t)$. For example, the $\tilde{p}_{x,y}^{m}(t)$ can be in [0,255] (e.g. choose hop count as the Rank metric), whereas $\triangle Q_{x,y}^{m}(t)$  may only between [0,10000]. To solve this problem, we can normalize the left and right operands as:

\begin{equation*}
w_{x,y}^{m}(t)=  \theta_{x}^m(t)\tilde{p}_{x,y}^{m}(t)
\\
-(1-\theta_{x}^m(t)) \triangle Q_{x,y}^{m}(t)\frac{c_{x,y}(t)}{c^{max}}
\label{eq:RankMRPLScaled}
\end{equation*}
\vspace{-0.6em}
\begin{eqnarray*}
&&\tilde{p}_{x,y}^{m}(t)= \frac{p_{x,y}^{m}(t)+Rank^m_y(t)}{\rm MaxRank}\\
&&\triangle Q_{x,y}^{m}(t) = \frac{{Q}^m_x(t)}{MaxQ_{x}^{m}} - \frac{Q^m_y(x,t)}{MaxQ_{y}^{m}(x)}
\end{eqnarray*}

Where  $\rm MaxRank\leq 2^{16}-1$ is the maximal rank value, depending on the Rank metric types (e.g. hop count, ETX). Here,  $2^{16}-1$ is  the maximum possible value of $\rm MaxRank$ specified in the RPL specifications. $w_{x,y}^{m}(t)$ in this case for any two pairs of nodes is always in $[-1,1]$ range.

\item\textbf{Routing and Data Forwarding.}
With the computed weights  $w_{x,y}^{m}(t),~\forall y \in \mathcal{N}_x(t)$, node $x$ will compute its \textit{potential} parent (i.e. next-hop destination) $y^*_m$ for DAG $m$:
\begin{equation}
y^*_m(x,t)=\arg\min_{y\in\mathcal{N}_x(t)}(w_{x,y}^m(t))
\end{equation}
Data packets for upward data traffic are forwarded to the potential parent $y^*$ if $w_{x,y^*_m}^{m}(t) > 0 \vee~ \triangle Q_{x,y^*_m}^{m}(t) > 0$. The actual data forward should be less than the current channel capacity and the number of data packets in its queue, i.e.

\begin{equation*}
  {f}_{x,y^{*}_m}^{m}(t)\leq\min({Q}^m_x(t), c_{x,y^*_m}(t))
\end{equation*}

\end{itemize}

BRPL has the following two special and interesting cases:
\begin{itemize}
\item \textit{Objective Function Optimality}: It is easy to observe that Eq. \ref{eq:RankMRPL} would be degraded to Eq. \ref{eq:rplRank}, when $\theta_{x}^m(t)=1$. Therefore, BRPL has identical performance to RPL when $\theta_{x}^m(t)=1~\forall x,y \in \mathcal{N}$ and both schemes greedily minimizes the routing objective specified by the OF.
\item \textit{High Throughput }: When $\theta_{x}^m(t)=0~\forall x,y \in \mathcal{N}$, BRPL would achieve high throughput, because its behavior ( Eq. \ref{eq:RankMRPL} ) will be similar to the  well-known throughput-optimal \textit{backpressure routing}~\cite{tassiulas1992stability}.
\end{itemize}

\subsection{QuickTheta}
\label{subsec:quickTheta}

The parameter $\theta_{x}^m(t)$ in Eq. \ref{eq:RankMRPL} can be configured directly by the users, but this means that users must have knowledge about the underline routing protocols and the current physical network infrastructure. This is not feasible in practice for dynamic IoT, especially in multi-user IoT systems. To abstract this complexity from the application layer, we propose \textit{QuickTheta}, a lightweight online algorithm that adaptively adjusts parameter $\theta_{x}^m(t)$, according to network traffic congestion levels, without having any assumption on the deployed applications or their expected traffic levels.

QuickTheta maintains a \textit{smooth queue length} $\bar{Q}_{x}^{m}(t)$, which is an \textit{Exponential Weighted Moving Average (EWMA)} of $Q_{y}^{m}(x,t)$, i.e.

\begin{equation*}
\bar{Q}_{y}^{m}(x,t)=
\begin{cases}
0&~t=1\\
\alpha\bar{Q}_{y}^{m}(x,t-1)+(1-\alpha)Q_{y}^{m}(x,t)&~t>1
\end{cases}\\
\end{equation*}

where $0 \leq \alpha \leq 1$ is the smoothing factor. Based on the current smooth queue length value, $\theta_{x}^{m}(t)$ is computed as:

\begin{equation}
\theta_{x}^{m}(t) = \beta_x(t) \left ( 1- \frac{1}{|\mathcal{N}_x(t)|+1}\sum_{y\in \mathcal{N}_x(t)\cup\{x\}}\frac{\bar{Q}_{y}^{m}(x,t)}{MaxQ_y^m(x)} \right )
\label{eq:brplQuickTheta}
\end{equation}

where the $\beta_x(t)$ parameter is for mobility-awareness, which is discussed in the next subsection. For now, assume $\beta_x(t)=1$. The ratio  ${\bar{Q}_{y}^{m}(x,t)}/{MaxQ_y^m(x)}$ is considered as a measurement of $y$'s \textit{local congestion level} for DAG $m$. This measurement relies on the usage share of the y's queue. The more packets are stored in the queues of the nodes $y \in \mathcal{N}_x(t)\cup\{x\}$, the closer $\theta_{x}^{m}(t)$ is to $0$. This results in pushing BRPL to increase network throughput based on Eq. \ref{eq:RankMRPL}.
The reasons behind the design choice of Eq. \ref{eq:brplQuickTheta} are presented in Section \ref{sec:rationale_QuickThetaBeta}.

\subsection{QuickBeta: Mobility Support}

\textit{QuickBeta} computes the $\beta_x(t)$ parameter in Eq. \ref{eq:brplQuickTheta} based on the mobility condition of the nodes. This is defined as follows:

\begin{equation}
\beta_x(t) = \frac{1}{\triangle t}\sum_{\tau=t-\triangle t}^{t-1}\frac{|\mathcal{N}_x(\tau) \cap \mathcal{N}_x(\tau+1)|}{\text{max}(|\mathcal{N}_x(\tau)\cup\mathcal{N}_x(\tau+1)|,1)}
\label{eq:brplQuickBeta}
\end{equation}

which observes the state changes of one-hop neighbor nodes for the node $x$ within the time window $[t-\triangle t,t)$. The more neighbors change their states (i.e. from online to offline or from offline to online), the closer $\beta_x(t)$ to 0 and the more node $x$ is seen as mobile.

For example, let node $x$ has three neighbors $\{A,B,C\}$ at slot $\tau$, and two existing neighbors $\{B,C\}$ leave and a new neighbor $D$ joins at slot $\tau+1$, i.e. $\mathcal{N}_x(\tau+1)=\{A,D\}$. We  compute the neighbor state changing rate for $x$ from slot $\tau$ to $\tau+1$ as

\begin{equation*}
\frac{|\mathcal{N}_x(\tau) \cap \mathcal{N}_x(\tau+1)|}{|\mathcal{N}_x(\tau)\cup\mathcal{N}_x(\tau+1)|}=\frac{|{\{A\}}|}{|\{A,B,C,D\}|}=0.25
\end{equation*}

From the Eq. \ref{eq:brplQuickTheta}, it is easy to see that the closer $\beta_x(t)$ is to 0 (i.e. the more node is mobile), the closer $\theta_{x}^{m}(t)$ is to 0 too. This causes BRPL to rely more on queue length differential for routing operations based on Eq. \ref{eq:RankMRPL}. In addition, the $\beta$ parameter in Eq. \ref{eq:brplQuickTheta} can be weighted to reduce or increase the sensitivity of mobility awareness in BRPL routing.

The rationale behind the design of QuickBeta algorithm is presented in Section \ref{sec:rationale_QuickThetaBeta}.

\section{Discussion and Analysis}
\label{sec:analysis}

\subsection{Design Principles of BRPL}
To provide a solution that is effective in IoT, the following factors have been considered in BRPL:

\subsubsection{Low Power and Lossy IoT} Similar to RPL, BRPL is designed mainly for LLNs using IPv6. The characteristics of LLNs are carefully considered in the proposed solution. In particular we focus on satisfying the limited power and processing resources. Control message overheads are intentionally kept to minimum. Only one new 6-byte field is added to RPL DIO control messages. Both QuickTheta and QuickBeta does not require statistical models or a learning/training phase to operate.

\subsubsection{Focus on many-to-one routing, yet any-to-any is still supported} BRPL focus mainly on the upward data traffic, which is predominant traffic patterns in LLNs. Here, the traffic is multipoint-to-point. As stated in \cite{ancillotti2014reliable}, other traffic patterns like unicast and point-to-multipoint are less frequent in LLNs. For these types of patterns, BRPL uses the two operation modes (\textit{non-storing} and \textit{storing}) that are originally defined by the specifications of RPL.

\subsubsection{Different Mobility Schemes} Although BRPL uses the mobility metric defined in Eq. \ref{eq:brplQuickBeta}, other mobility metrics can be easily integrated with BRPL. This includes, but not limited to,  rendezvous-based \cite{xing2008rendezvous}, trajectory-based \cite{chon2011mobility} and social-aware \cite{yang2013selfish} mobility metrics.

\subsection{Design Rationale for QuickTheta and QuickBeta}
\label{sec:rationale_QuickThetaBeta}
When we designed  QuickTheta and QuickBeta, we considered a wide range of factors to ensure having an implementable and practical solution in LLNs. The following list briefly highlights three advantages of our algorithm design:

\textit{Single-layer Dependency}: Both QuickTheta and QuickBeta are self-contained and relies on one network layer only, the routing layer. This is desirable design property to have. Making QuickTheta and QuickBeta independent from the other network layers not only enables seamless integration with the OSI model, but also ensures usability under diverse communication systems. For example, utilizing a custom MAC layer does not cause any interoperability issue for the two algorithms.

\textit{Simplicity}: The choice of Eq. \ref{eq:brplQuickTheta} and \ref{eq:brplQuickBeta} is suitable for nodes with limited amount of resources. Both algorithms are simple to compute, and they do not rely on sophisticated statistical models. There is no need to perform resource-hungry data analysis operations here, nor to keep extensive historical data from multiple network layers. The QuickBeta algorithm only needs to retain the value of $\mathcal{N}_x(t)$ for the time slots between ($t - \triangle t$) and $t$. The $\bar{Q}_{y}^{m}(x,t)$ for all the nodes in $y \in \mathcal{N}_x(t)$ in the QuickTheta algorithm can be maintained in a vector of  $\mathcal{N}_x(t) + 1$ elements.

\textit{User Abstraction and Self Parameter Tuning}: It is easy to observe that both Eq. \ref{eq:brplQuickTheta} and \ref{eq:brplQuickBeta} do not incorporate a direct feedback from network users, instead they rely on neighbor table and network traffic congestion levels to adjust their performance. This is intentional and by design because it makes BRPL more suitable for multi-user networks, specially when a network has a DAG serving multiple greedy users. Even when user applications are highly dynamic and unpredictable, users are not required to tune QuickTheta and QuickBeta at runtime. This design also offers an effective solution to overcome user \textit{selfishness} as a user cannot mislead the QuickTheta and QuickBeta algorithms to gain a better position in the network.

\subsection{RPL Backward Compatibility Support}
\begin{figure}
 \centering
   \includegraphics[trim=60 200 300 60,clip=true,angle=0,width=0.45\textwidth] {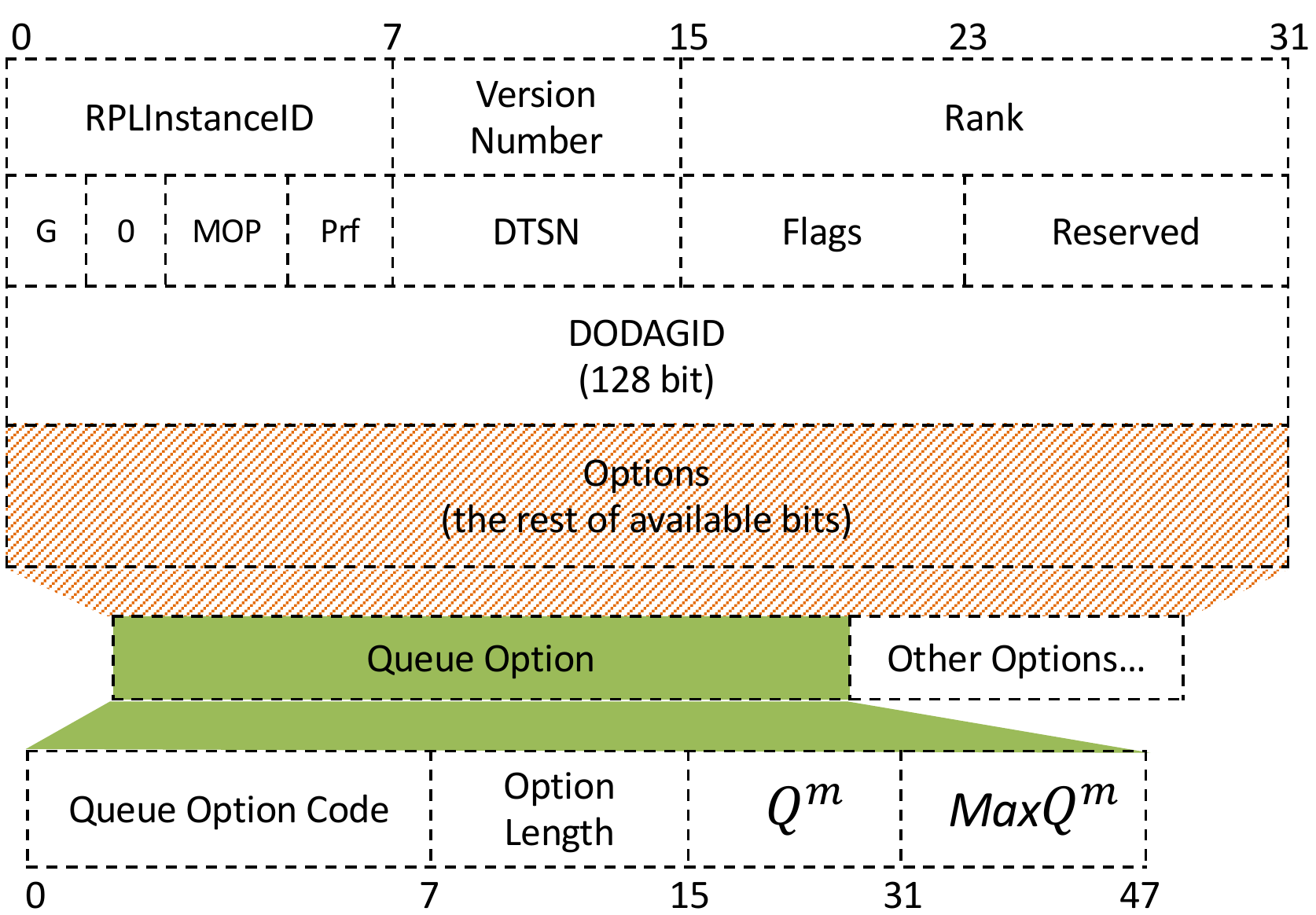} 
   \vspace{-1em}
  \caption{DIO message structure in BRPL}
  \label{fig:DIOStrcture_BRPL}
  \vspace{-0.5em}
\end{figure}

To ensure backward compatibility with RPL, BRPL does not introduce any new control message types but rather reuses the control message structures defined in RPL. Node $x \in \mathcal{N}$ broadcasts its queue length $Q_x^{m}$ to $\mathcal{N}_x(t)$ at time slot $t$ via DIO messages. BRPL introduces a new standard ICMP option named \code{Queue Option} as shown in Fig. \ref{fig:DIOStrcture_BRPL}. The payload of this new option has a length of 4 bytes stored in big endian order.

Maintaining interoperable communications between BRPL and non-BRPL nodes is a key element towards supporting hybrid networks. When a node running RPL processes an incoming DIO message with the Queue Option, it considers the option code to be unknown. The node will then ignore this ICMP option and process the rest of DIO message safely. This ensures a transparent message structure between BRPL and non-BRPL nodes, and allows nodes to seamlessly join hybrid networks without experiencing compatibility-related issues.

\begin{figure}
 \centering
   \includegraphics[trim=40 210 220 20,clip=true,angle=270,width=0.45\textwidth] {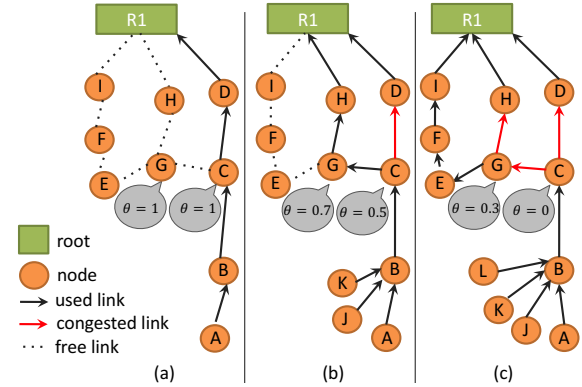} 
  \caption{An illustration demonstrates how BRPL supports large-scale networks. Assume nodes can concurrently listen to multiple radio channels. In (a), the network is stable and the path B-C-D-R1 handles the data traffic. Both BRPL and RPL produce the same DAG. In (b) and (c), new nodes join the network. BRPL automatically adjusts $\theta$ to utilize additional network resources.}
  \label{fig:BRPLScaleCase}
  \vspace{-1em}
\end{figure}

\vspace{0.5em}
\subsection{Support for Large-scale Networks}

The high adaptability in BRPL also enables deploying large-scale LLNs. When new nodes join the network, they may introduce extra traffic to the network causing BRPL to adjust the $\theta$ parameter accordingly. Unlike RPL, when network bottleneck appears in the network, BRPL tries to solve it by allocating more resources for the incoming data traffic. Similarly, when nodes leave the network and traffic level becomes lower, BRPL deallocates resources that are  no longer required.

Fig. \ref{fig:BRPLScaleCase} presents an example illustrating how BRPL naturally supports large-scale networks. Let assume that R1, the root of the network, offers multiple communication channels. In (a), the path B-C-D-R1 handles all the data traffic coming from node A. Both RPL and BRPL have the exact same DAG since $\theta$ is close 1 when the traffic load is light. In (b), node K and J join the network. Now the path B-C-D-R1 becomes congested and cannot handle all the incoming traffic. Unlike RPL, BRPL observes the traffic congestion in the DAG, and diverts some of the traffic through the path B-C-G-H-R1. In (c), one more node joins the network resulting in further traffic congestion. Again BRPL adjusts the routing and allocates more resources trying to eliminate the traffic congestion. When nodes L, K, J leaves the network, BRPL deallocates the extra resources utilized in (c) and (b) causing the DAG to return back to the DAG showing in (a).

\section{Experiments}
\label{sec:evaluation}

The practical performance of BRPL has been compared to RPL and backpressure routing. This section firstly describes the general configuration settings of our experiments. Then, it shows BRPL performance in static networks using real-life IoT testbed, and mobile networks using a network simulator with \textcolor{black}{a realistic} indoor mobility model.

\begin{figure}[t!]
\centering
     \includegraphics[trim=120 120 120 240,clip=true, angle=270,width=0.35\textwidth]{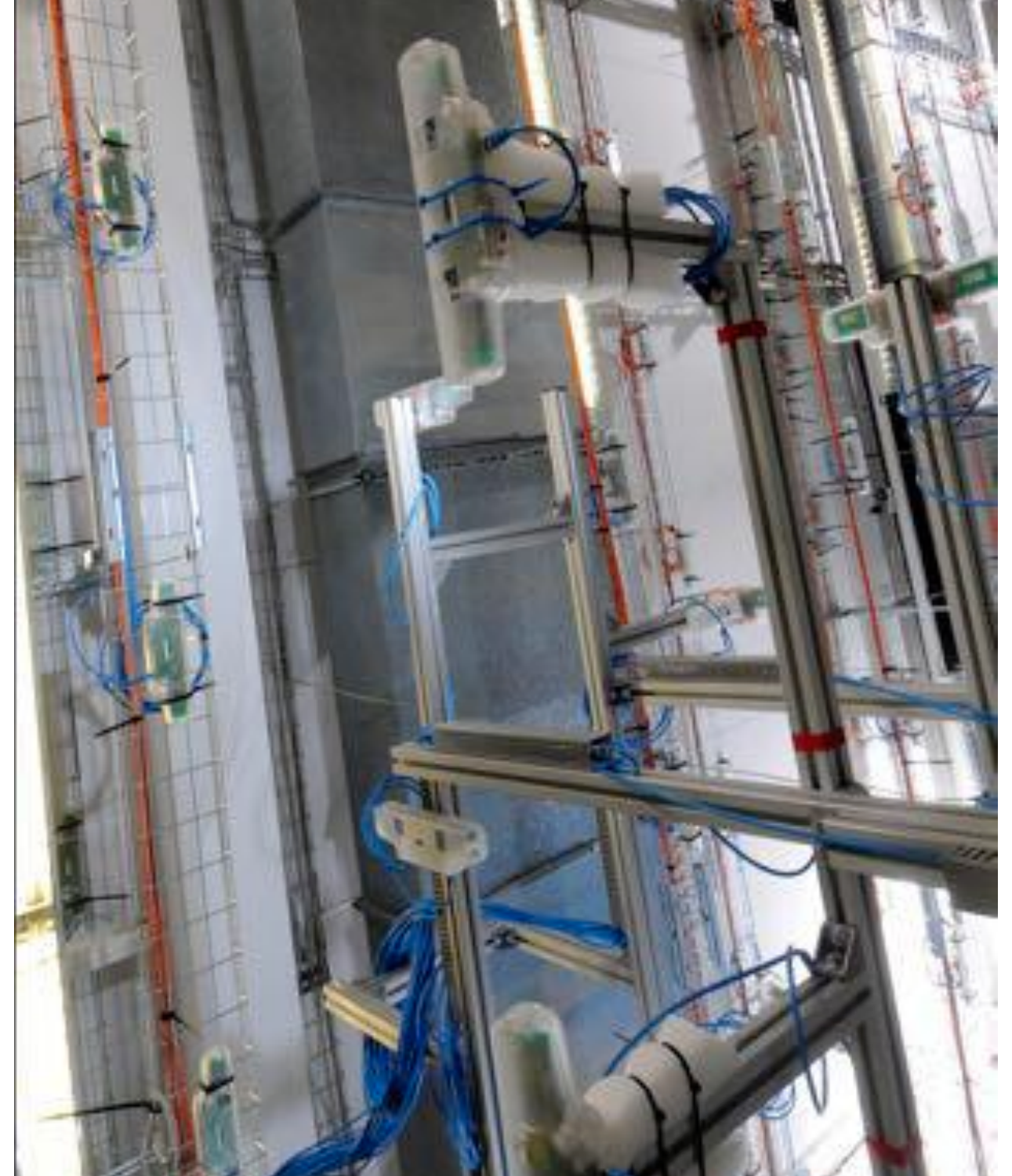} 
    \caption{Node deployment in the FIT IoT-LAB testbed.}
    \label{fig:testbed_networktopology}
	\vspace{-1em}
\end{figure}

\subsection{Implementation and Configuration}
\label{sub:configEvaluation}
BRPL is implemented on top of Contiki OS \cite{contiki}, an open source operating system designed for systems with limited resources including, but not limited to, LLNs and IoT. Out of the box, Contiki provides a C implementation for the IPv6 stack (uIPv6) and RPL. This section summarizes some key modifications that we have done in Contiki before running the experiments.

For the MAC layer, a CSMA/CA driver combining with MiCMAC \cite{micmac} was used. MiCMAC is mainly employed to perform channel hopping. The maximum number of channels that a node can use is 4 and the maximum channel cycle listening time is 80ms. The neighbor table size is 50 records. The radio duty cycling is disabled during all experiments. The maximum transmission attempts to re-send a packet is set to 5.

In regards to the IPv6 layer, we set the \code{UIP\_CONF\_ND6\_REACHABLE\_TIME} parameter to 5000 and the \code{UIP\_CONF\_ND6\_MAX\_UNICAST\_SOLICIT} to 65535 to increase the effectiveness of IPv6 Neighbor Discovery service.  The packet reassembly service is disabled and the \code{UIP\_CONF\_IGNORE\_TTL} is set to zero to ignore the TTL flag in the packet headers. The \code{HC6} SICSlowpan header compression is used in all our experiments.

\begin{table}
\centering
\caption{Summary of Evaluation Parameter Settings}
\label{brpl_evalutionSettingTable}
\vspace{-1em}
\begin{tabular}{|l|c|c|c|}
\hline
\textbf{Method} & \textbf{Testbed}& \multicolumn{2}{c|}{\textbf{Simulation}}                                               \\ \cline{2-4}
\textbf{} & \textbf{IoT-LAB}& \textbf{Factory} & \textbf{QuickTheta} \\ \hline
\textbf{OF} & ETX & ETX & Custom \\ \hline
\textbf{Platform}& \begin{tabular}[c]{@{}c@{}}Contiki/\\ M3 ARM\end{tabular} & \begin{tabular}[c]{@{}c@{}}Contiki/\\ Cooja\end{tabular} & \begin{tabular}[c]{@{}c@{}}Contiki/\\ Cooja\end{tabular} \\ \hline
\textbf{MAC} & \begin{tabular}[c]{@{}c@{}}CSMA/CA+\\ MiCMAC\end{tabular} & \begin{tabular}[c]{@{}c@{}}CSMA/CA+\\ MiCMAC\end{tabular} & \begin{tabular}[c]{@{}c@{}}CSMA/CA+\\ MiCMAC\end{tabular} \\ \hline
\textbf{Mobility} & - & \begin{tabular}[c]{@{}c@{}}Factory Indoor \\ Mobility\end{tabular} & -\\ \hline
\textbf{Transport}& UDP/IPv6 & UDP/IPv6 & UDP/IPv6\\ \hline
\textbf{Packet Size} & 160 bytes & 160 bytes & 160 bytes\\ \hline
\textbf{Queue Size} & 150 packets & 250 packets & 250 packets \\ \hline
\textbf{\begin{tabular}[c]{@{}l@{}}Transmission \\ Power\end{tabular}} & -17 dBm                                                   & 0 dbm & 0 dbm \\ \hline
\textbf{\begin{tabular}[c]{@{}l@{}}Transmission \\ Range\end{tabular}} & -                                                         & 30 meter & 50 meter \\ \hline
\textbf{\begin{tabular}[c]{@{}l@{}}Network \\ Scale\end{tabular}} & 100 nodes                                                 & 130 nodes & 100 nodes\\ \hline
\end{tabular}
\end{table}

For the routing layer, ETX OF has been utilized with the default parameter settings. We also enable the \code{RPL\_MOP\_NO\_DOWNWARD\_ROUTES} option since downward routing is not used in our experiments. \textcolor{black}{Time slot duration is assumed to be one second. \code{DIO\_INTERVAL\_MIN} and \code{DIO\_INTERVAL\_DOUBLINGS} were adjusted  to broadcast routing metadata every 512 to 1024ms period. When the network is highly dynamic (i.e. time-varying traffic and/or topology), the Trickle algorithm pushes the DIO broadcasting rate towards every 512ms. On the other hand, Trickle adjusts the broadcasting rate to 1024ms when weight stability is observed in one-hop neighbors.}

The ``Queue Manager'' component has been implemented, and integrated with the IPv6 stack in Contiki. The Queue Manager uses Last-In-First-Out (LIFO) scheduling. When a node with full data queue receives a data packet, it drops the newly received packet. In addition, to make RPL more fair and reduce its packet loss, data queues has been added to RPL to buffer the incoming packets. However unlike BRPL, data queues in RPL are not considered during routing operations. The queue option code in DIO messages is set as `0xCE', which does not interfere with existing RPL option codes in Contiki.

The \textcolor{black}{rest} of BRPL components shown in Fig. \ref{fig:BRPL_ARCH} has been also implemented successfully in Contiki. In the experiments, BRPL uses the same OF of RPL with the same settings. The neighbor table size of both BRPL and RPL is 50.  In addition, we implemented the well-known backpressure routing \cite{tassiulas1992stability} on top of the IPv6 stack, but without its scheduling policy. \textcolor{black}{All three routing protocols have identical queue settings.}

The performance of BRPL, RPL and backpressure routing has been examined in static and mobile networks. The length of each experiment is 4 hours. The application layer generates UDP packets on a regular basis and passes them to the IPv6 layer. The size of a data packet is 160 bytes (8 bytes for payload, the rest is used for IPv6 header). Table \ref{brpl_evalutionSettingTable} highlights key \textcolor{black}{settings in our experiments}.

\subsection{Methodology}

Both testbed experiments and simulations are employed to evaluate the performance of BRPL.
\subsubsection{Testbed Experiments.} We used 100 nodes (M3 Open Nodes) from the Grenoble site offered by the FIT IoT-LAB testbed \cite{iotlab}, shown in Fig. \ref{fig:testbed_networktopology}. The network had 5 roots and 95 sensor nodes generating UDP packets on predefined time interval. Each M3 open node has a ARM Cortex M3 micro-controller, a 64 kB RAM, a IEEE 802.15.4 radio AT86RF231, several types of sensors, and a rechargeable 3.7 V LiPo battery. To ensure multi-hop mash topology is formed, we set transmission power to -17 dBm. The $MaxQ_x^m$ is set to 150 for all nodes.

\subsubsection{Cooja Simulations based on Cloud.} To evaluate how BRPL can extends RPL to support mobile IoT scenarios, we used Cooja - the network simulator of Contiki. We set the $MaxQ_x^m$ to 250 packets. The transmission range is adjusted to 30m to simulate indoor radio limitations. We noticed that Cooja has a poor performance in a 130-node network. To speed up, we run all our simulations on a 18-server cluster provided by the Imperial College London's private Cloud. Each server runs 14.04 Ubuntu LTS with 4 GB of RAM. Git and Puppet, an open-source configuration management tool, are utilized in order to run multiple simulations in parallel with different configuration settings.

\begin{figure}
\centering
	\subfigure[packet loss.]{
		\includegraphics[trim=0 0 0 0,clip=true,angle=0,width=0.48\textwidth] {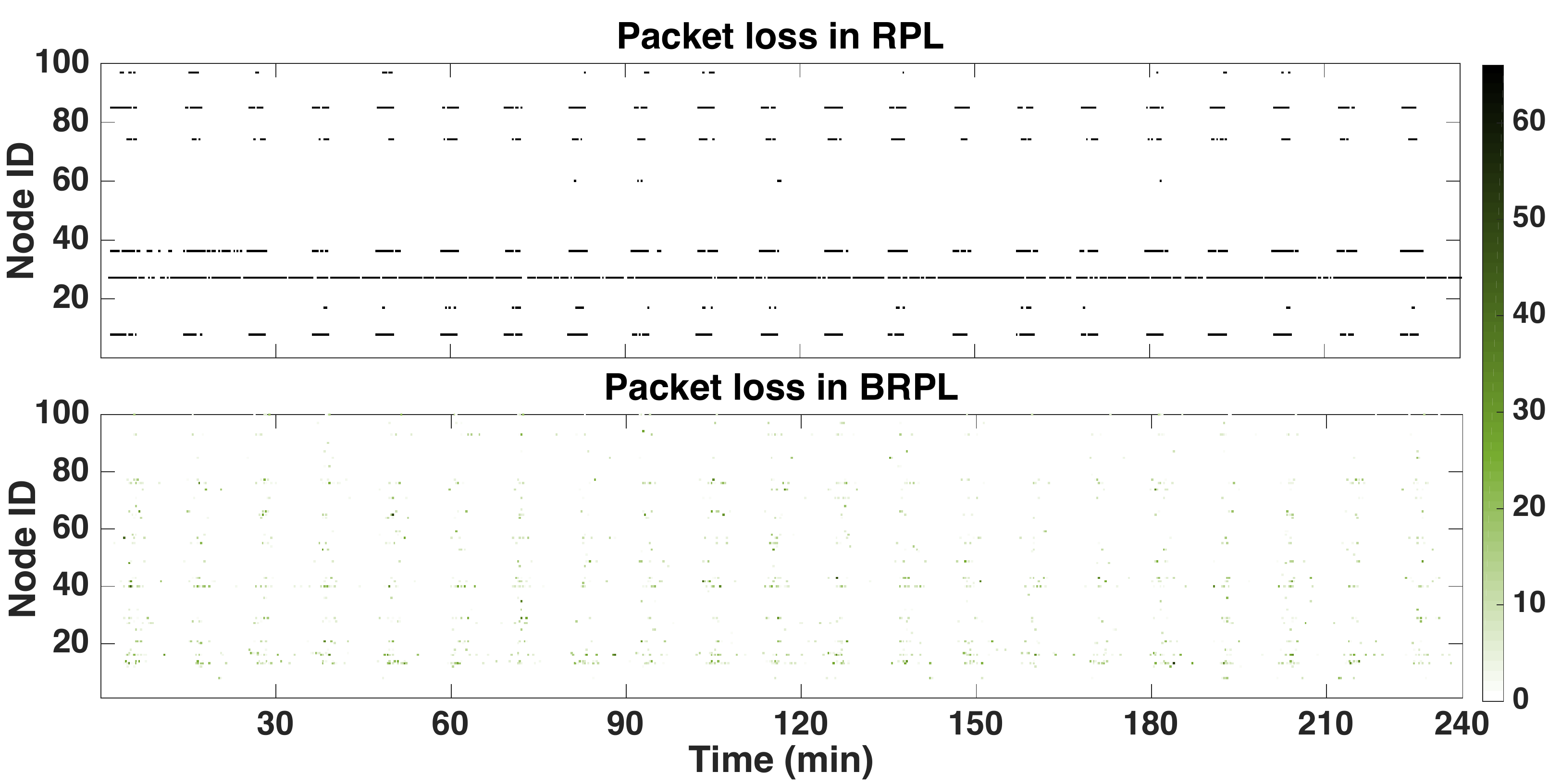} 
		\label{fig:testbed_varyingBRPLvsRPL}
	}\vspace{-0.5em}
	\subfigure[$\theta$ parameter at runtime.]{
		\includegraphics[trim=0 0 0 0,clip=true,angle=0,width=0.49\textwidth] {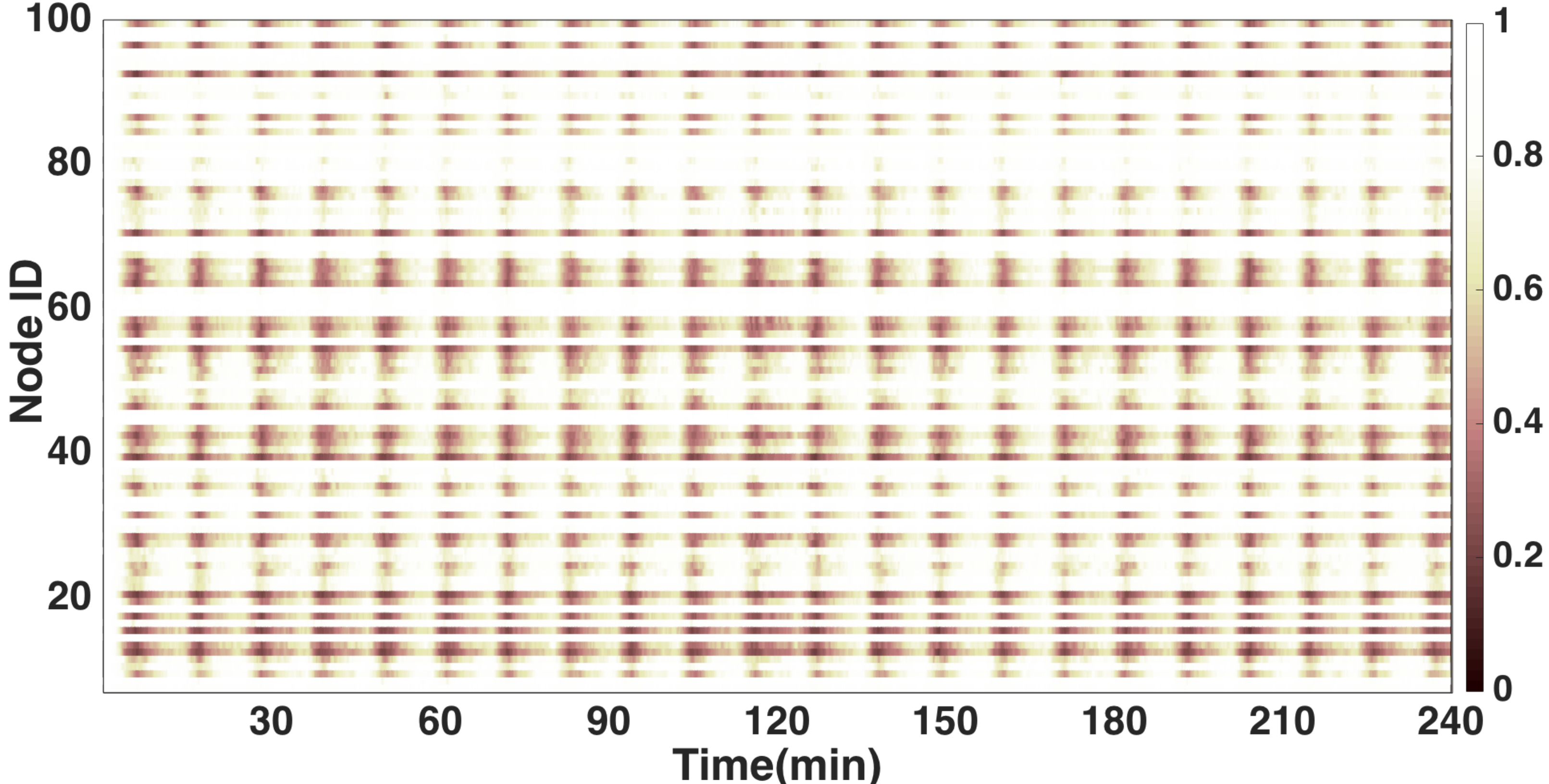} 
		\label{fig:testbed_varyingTheta} %
	}
    \vspace{-1em}
	\caption{Testbed results when the network has dynamic data traffic. Every 10 minutes, the application layer increases its sensing rate to generate a traffic burst last 3 minutes. (a) shows the packet loss for BRPL and RPL. (b) shows the $\theta$ parameter and its adaptivity at runtime.}
	\label{fig:testbed_varying}
\vspace{-1.0em}
\end{figure}

\subsection{Testbed Experiments}

\begin{figure*}[h!]
\centering
	\subfigure[packet loss.]{
		\includegraphics[trim=20 0 0 0,clip=true,angle=0,width=0.33\textwidth] {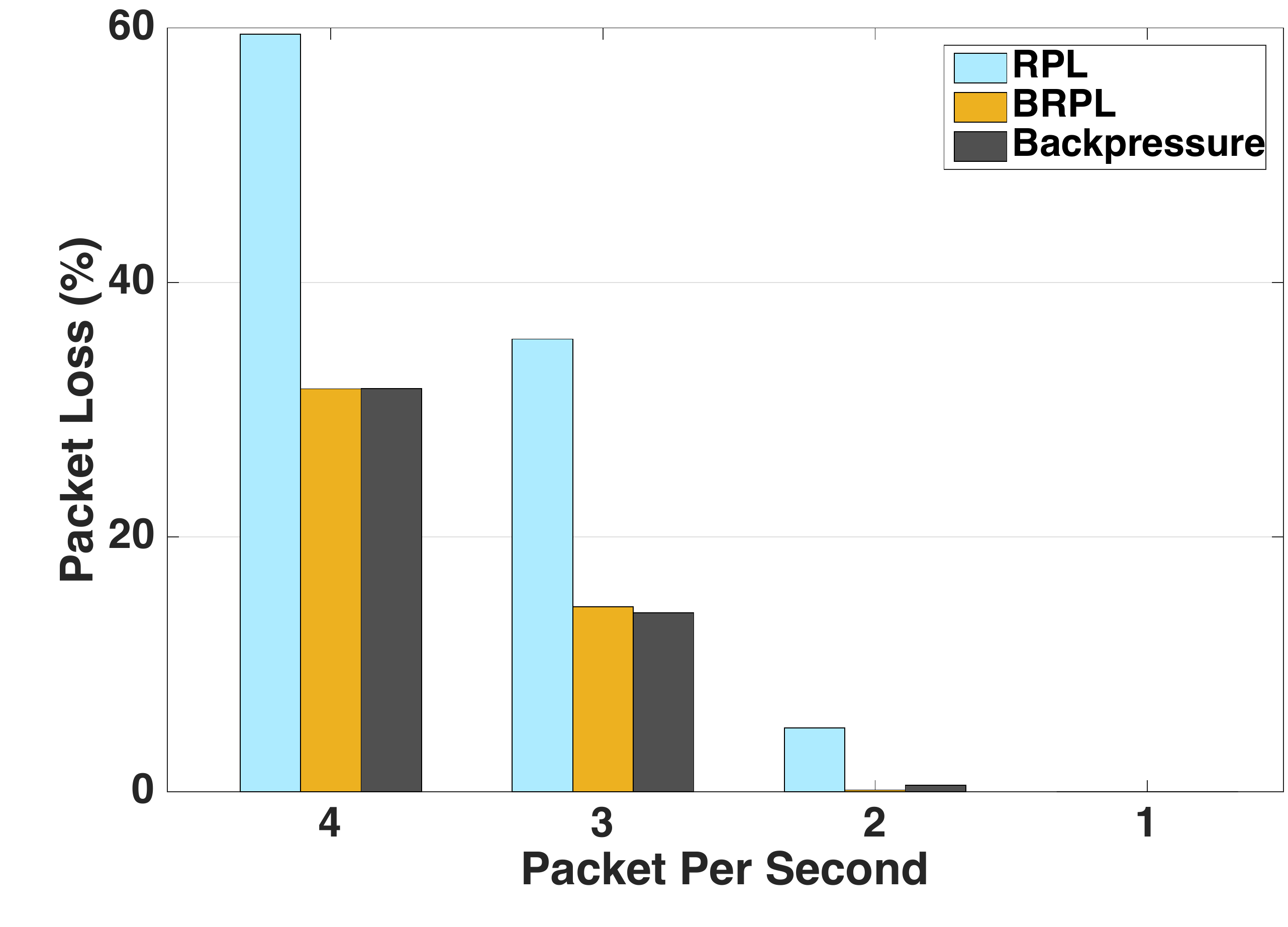} 
		\label{fig:testbed_packetLoss}
	}~
	\subfigure[end-to-end delay.]{
		\includegraphics[trim=20 0 0 0,clip=true,angle=0,width=0.33\textwidth] {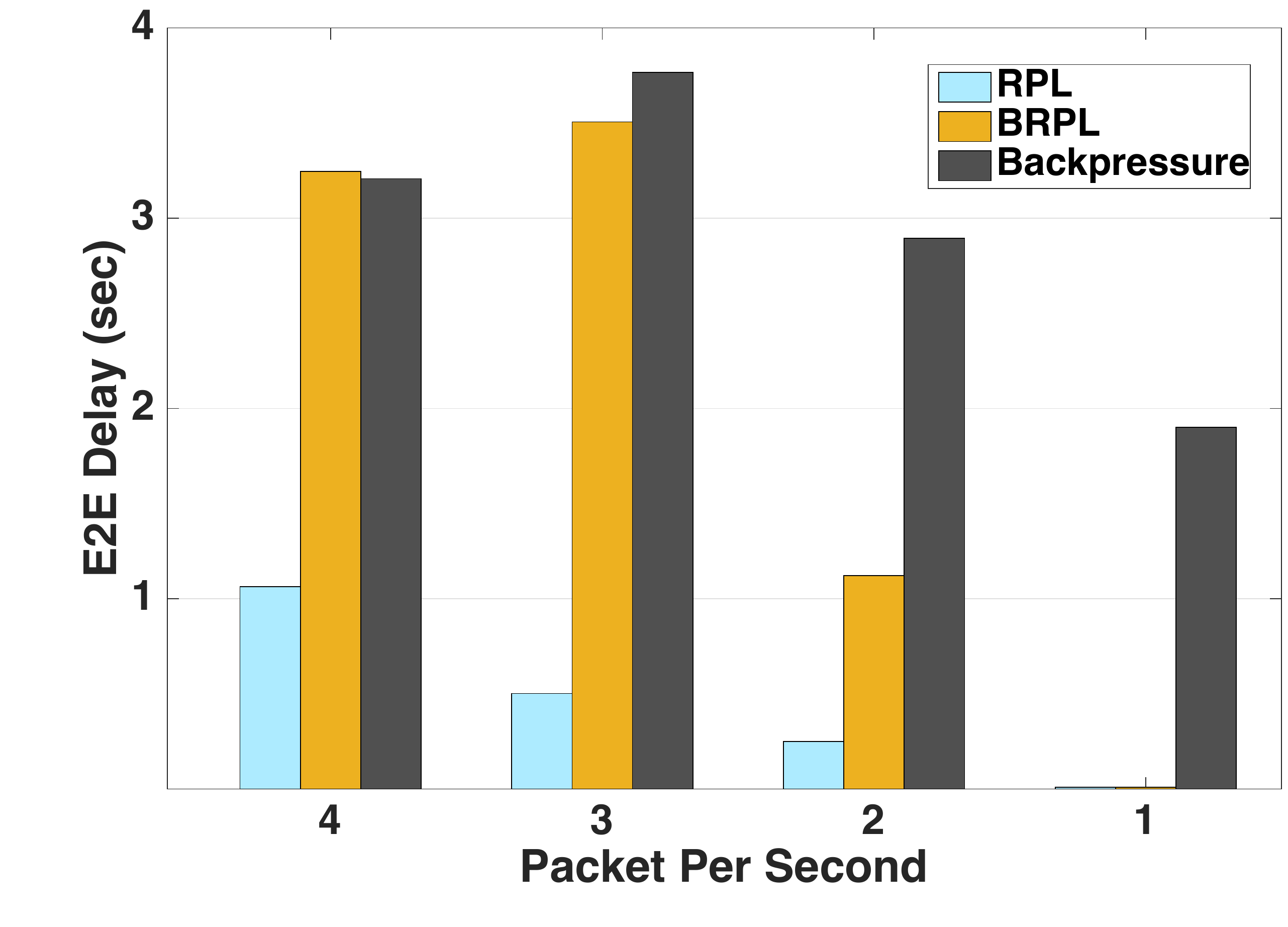} 
		\label{fig:testbed_delay}
	}~
	\subfigure[communication overhead (in $10^6$  packets).]{
		\includegraphics[trim=20 0 0 0,clip=true,angle=0,width=0.33\textwidth] {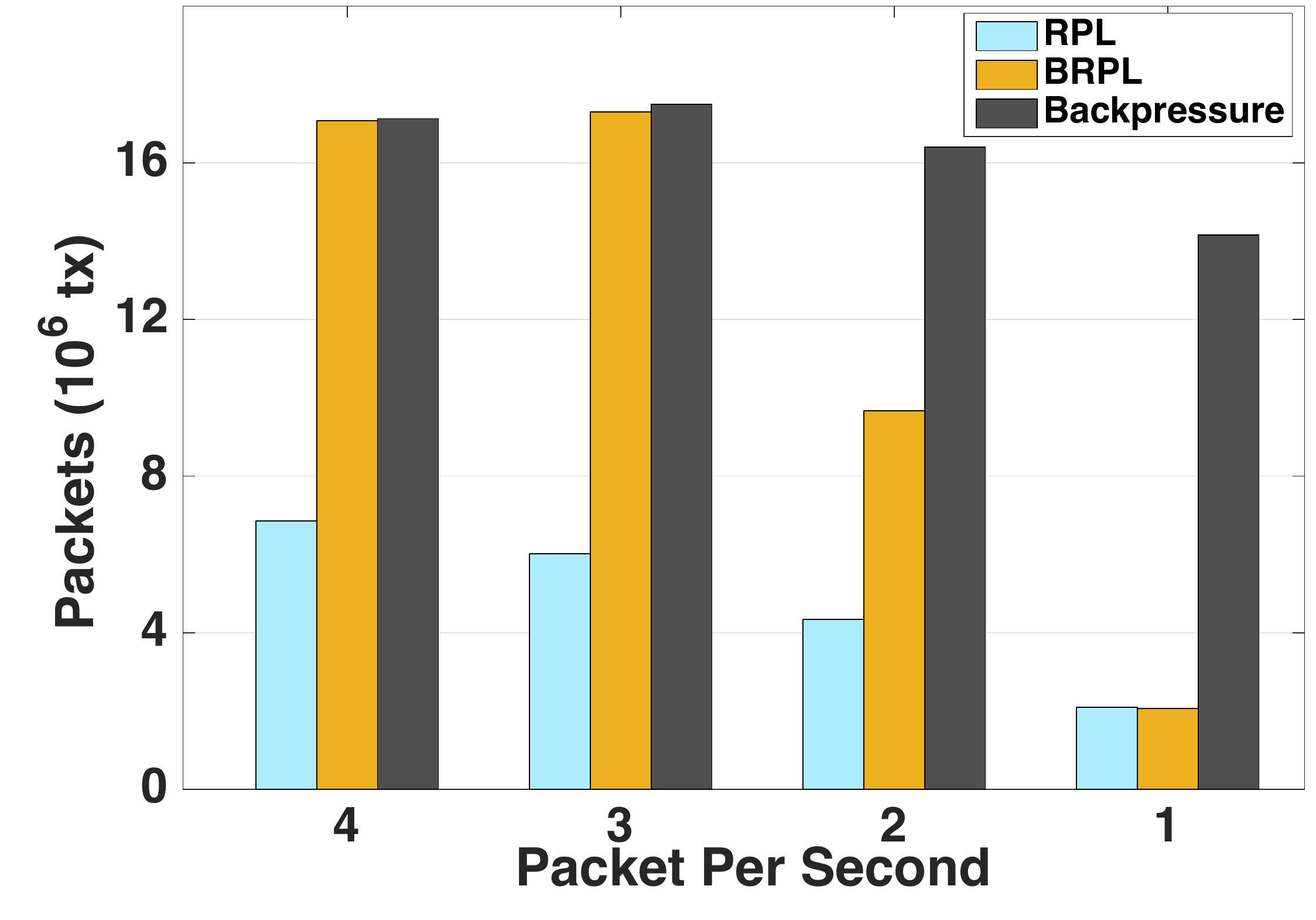} 
		\label{fig:testbed_control}
	}
    \vspace{-1em}

	\caption{The performance of BRPL, RPL and backpressure under different traffic loads in the testbed. RPL has the highest packet loss.  BRPL and backpressure routing uses suboptimal paths when the traffic load is high (i.e when data rate is 4, 3 or 2 PPS) which results in a significant reduction in packet loss with an increase in end-to-end packet delay and communication overhead. When the traffic load is light (e.g. 1 PPS), both BRPL and RPL have similar performance, whereas backpressure routing still suffers from large delay and communication overhead.}
    \vspace{-1em}
\end{figure*}

\begin{figure*}[h!]
\centering
	\subfigure[packet loss.]{
		\includegraphics[trim=20 0 0 0,clip=true,angle=0,width=0.33\textwidth] {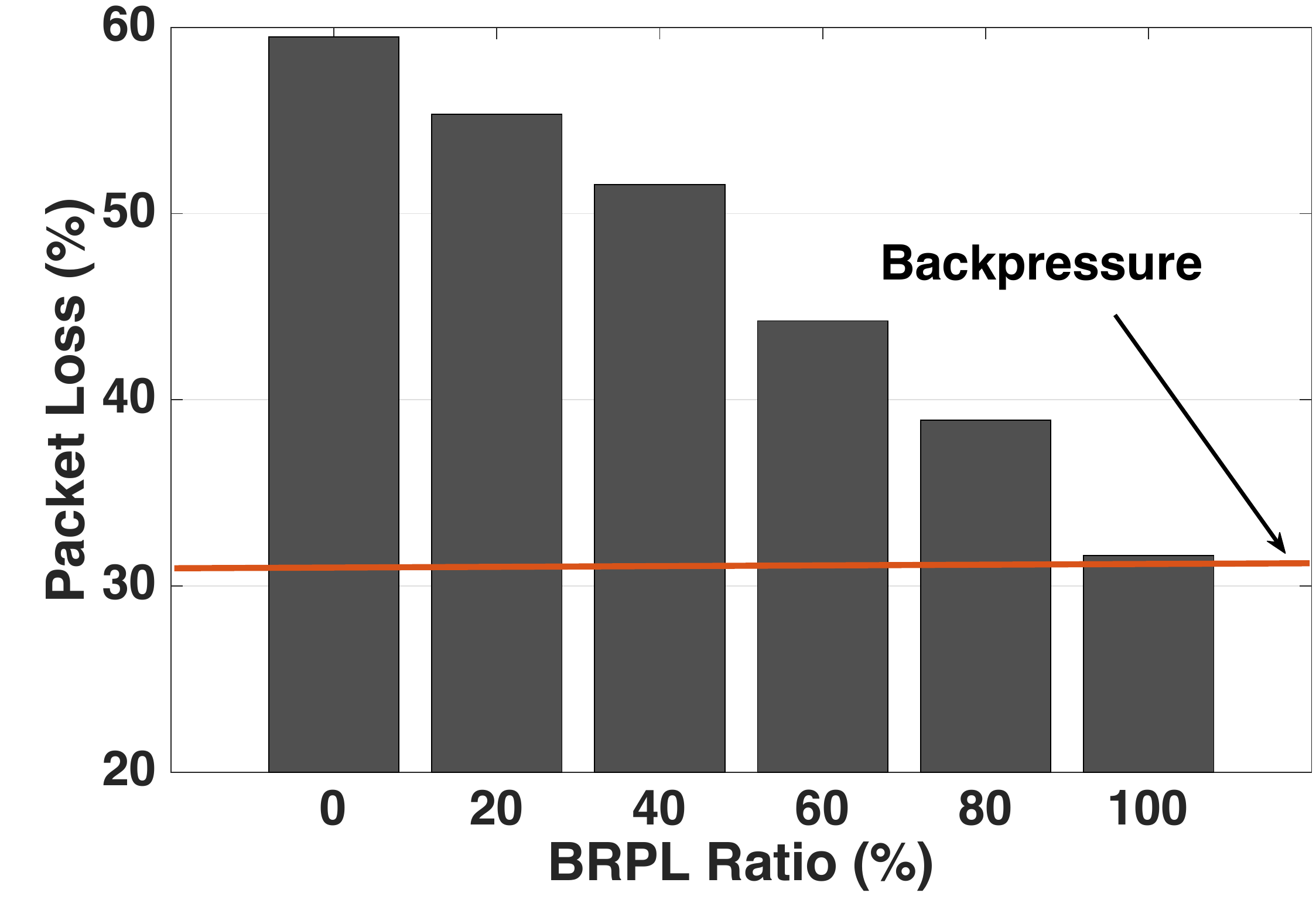} 
		\label{fig:testbed_hybridPacketLoss}
	}~
	\subfigure[end-to-end delay.]{
		\includegraphics[trim=20 0 0 0,clip=true,angle=0,width=0.33\textwidth] {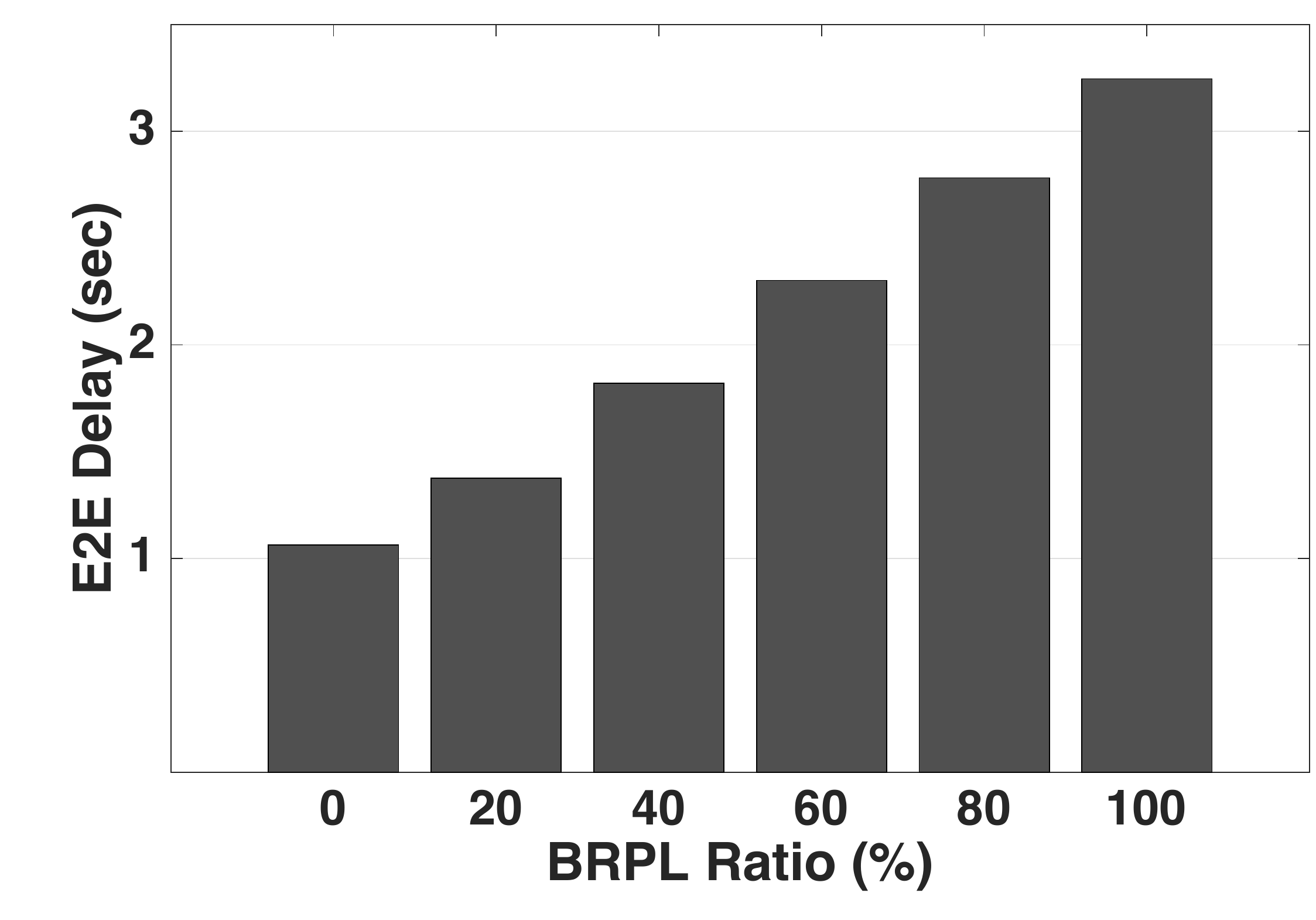} 
		\label{fig:testbed_hybridDelay}
	}~
	\subfigure[communication overhead (in $10^6$  packets).]{
		\includegraphics[trim=20 0 0 0,clip=true,angle=0,width=0.33\textwidth] {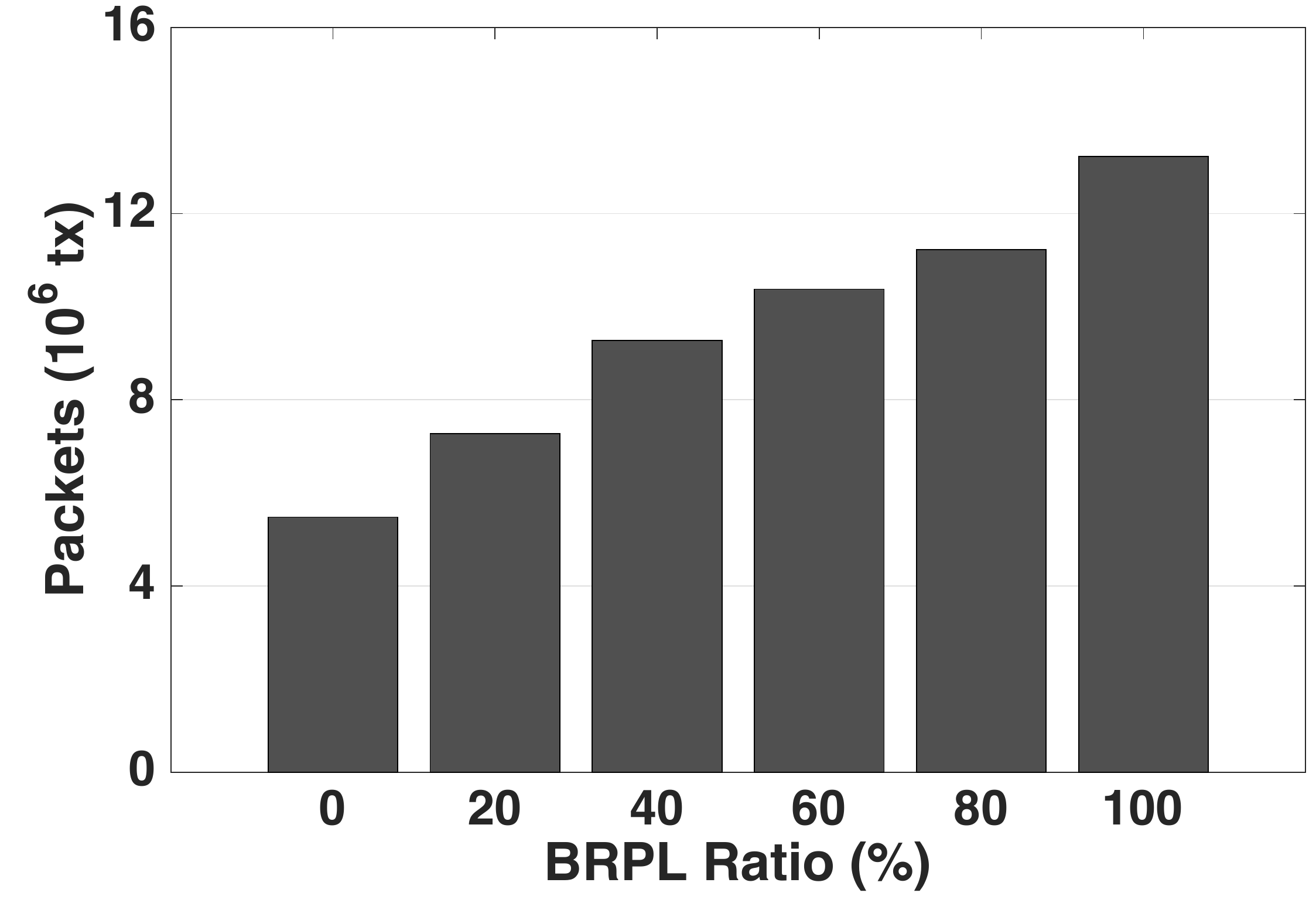} 
		\label{fig:testbed_hybridControl}
	}
    \vspace{-1em}
	\caption{The performance of a hybrid network with various BRPL node deployments. The data rate is fixed to 4 PPS. As seen, the more BRPL nodes deploy in the network, the lower packet loss is expected. This is accommodated with higher delay and communication overhead as BRPL nodes may use suboptimal OF paths to improve the throughput.}
    \vspace{-1em}
\end{figure*}

Three sets of testbed experiments were constructed to compare practical performance of BRPL, RPL and backpressure routing based on  a 100-node network in \textcolor{black}{the FIT IoT-LAB testbed} with settings described in Section \ref{sub:configEvaluation}.

\subsubsection{Adaptivity to Dynamic Traffic Load} This set of experiments examines the performance of BRPL and RPL in a network with time-varying traffic loads. A simple application has been developed to generate data packets at rate 1 Packet Per Second (PPS). The application has an internal timer that is triggered every 10 minutes to change the packet rate to 4 PPS for three minutes. This simulates a traffic burst condition in the network. Fig. \ref{fig:testbed_varyingBRPLvsRPL} shows packet loss for BRPL and RPL as a function of time. RPL drops around 154K packets mainly when the network encounters a traffic burst. On the other hand, BRPL adapts to the traffic dynamics, thanks to the QuickTheta algorithm, and utilizes resources to handle the traffic burst. Fig. \ref{fig:testbed_varyingTheta} shows how the QuickTheta algorithm adjusts the $\theta$ parameter at runtime according to the traffic congestion levels. This runtime adaptivity results in BRPL having more than 4.5 times less packet \textcolor{black}{loss} than RPL.

\subsubsection{Throughput Study} In this set of experiments, the data sensing rate is fixed over time. Fig. \ref{fig:testbed_packetLoss} shows the packet loss BRPL, RPL and backpressure routing under different data rates. RPL has the highest packet loss as it tends to always choose the optimal OF path. To enhance the throughput of the network, BRPL utilizes suboptimal paths when the network \textcolor{black}{has} high data traffic (when the data rate is 4, 3, or 2 PPS). This results in around 100\% reduction in the packet loss which is accommodated with higher end-to-end packet delay and communication overhead\footnote{ \textcolor{black}{Please note that communication overhead reflects not only network congestion levels, but also the energy consumption of nodes. This is because it is well recognized that  packet transmissions are the major energy consumer of routing protocols~\cite{pantazis2013energy,yang2016distributed}. }}, as seen in Fig. \ref{fig:testbed_delay} and \ref{fig:testbed_control} respectively. It is important to note that both backpressure routing and BRPL have similar performance in terms of throughput. However, when the traffic load is light (e.g. 1 PPS), backpressure routing suffers from large high end-to-end packet delays. This is because backpressure routing relies on congestion gradients to route the packets. The data traffic in some settings was not enough to establish stable congestion gradients. This results in network frequently experiencing routing loops. BRPL, on the other hand, relies on the OF to route the packets in this case, as the $\theta$ in this case is close to 1. Data is forwarded via the optimal paths defined by the OF. This also explains why the performance of BRPL and RPL is very similar when the data rate is 1 PPS.

\subsubsection{Support for Hybrid Networks.} a multi-hop network where are all the nodes running RPL has been constructed. The data rate in this experiment is set to 4 PPS to simulate that the network is facing relatively high traffic load. Then we randomly replaced 20 nodes of RPL with BRPL nodes. The results were gathered, and then the same process was repeated until \textcolor{black}{all} the nodes run BRPL. From these experiments, the following conclusion can be realize: first, BRPL is backward compatible with RPL. Nodes running RPL or BRPL can operate together in the same hybrid network. Interoperability for data and control messages between RPL and BRPL nodes is successfully verified. Second, the more BRPL nodes deployed in the network, the less packet loss is observed as shown in Fig. \ref{fig:testbed_hybridPacketLoss}. Because the traffic level of the network in the experiments is considerably high, BRPL nodes tends to utilizes all the possible network capacity to increase the throughput and reduce the packet loss as much as possible. This is accommodated with an increase in average end-to-end delay and communication overhead as shown in Fig. \ref{fig:testbed_hybridDelay} and Fig. \ref{fig:testbed_hybridControl} respectively.

\subsection{Cooja-based Simulations}

Cooja network simulator is used to provide more insights about the performance of BRPL under other network conditions. These simulations firstly show BRPL performance for a network following an indoor mobility model. Then, we examine how QuickTheta algorithm achieves the \textcolor{black}{practical} balance between throughput and minimizing OF.

\begin{figure*}[t]
 \centering
   \includegraphics[trim=0 0 0 0,clip=true,angle=0,width=0.64\textwidth] {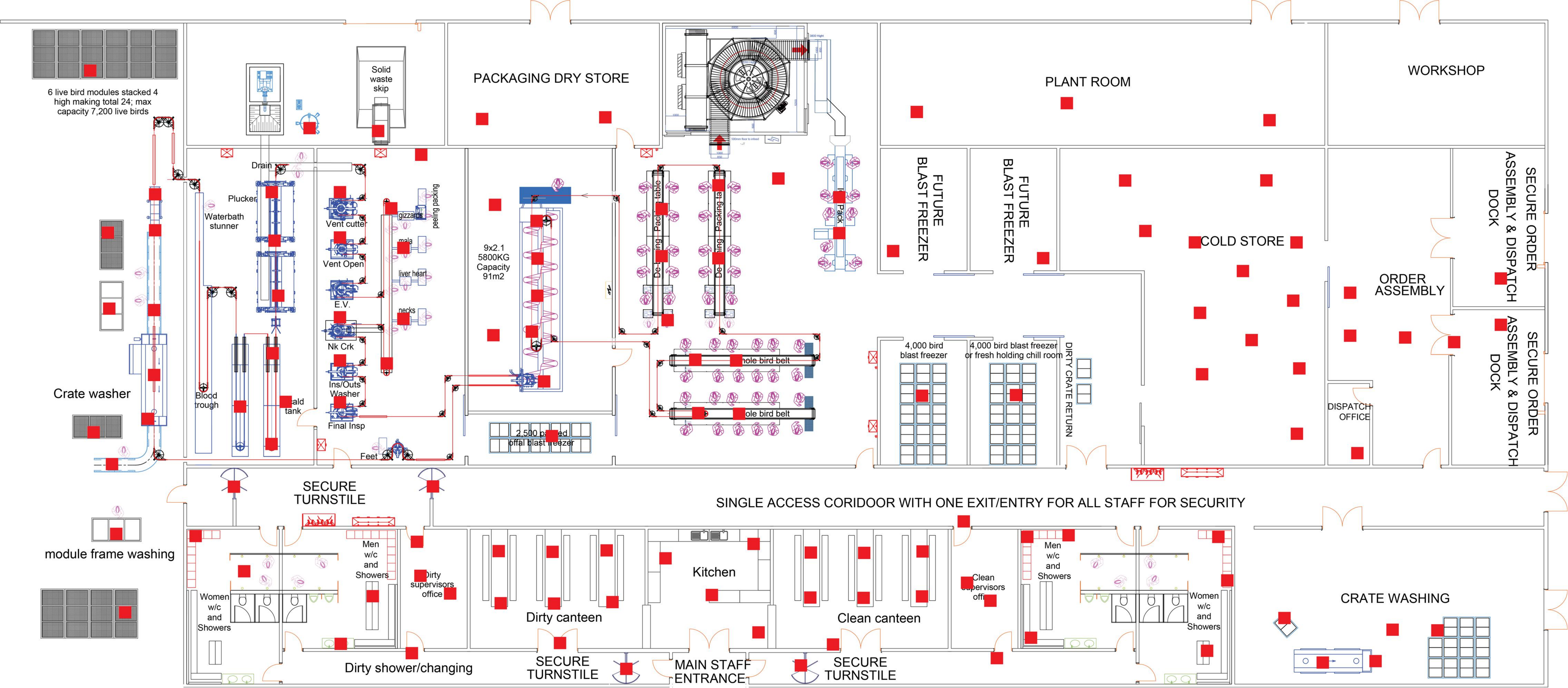} 
  \caption{A poultry processing factory layout obtained from \cite{poultryfactory}. The red squares denote the locations of the deployed sensor nodes.}
  \label{fig:factorySensors}
  \vspace{-1em}
\end{figure*}

\begin{figure*}[t]
 \centering
   \includegraphics[trim=0px 0 0px 0,clip=true,angle=0,width=0.64\textwidth] {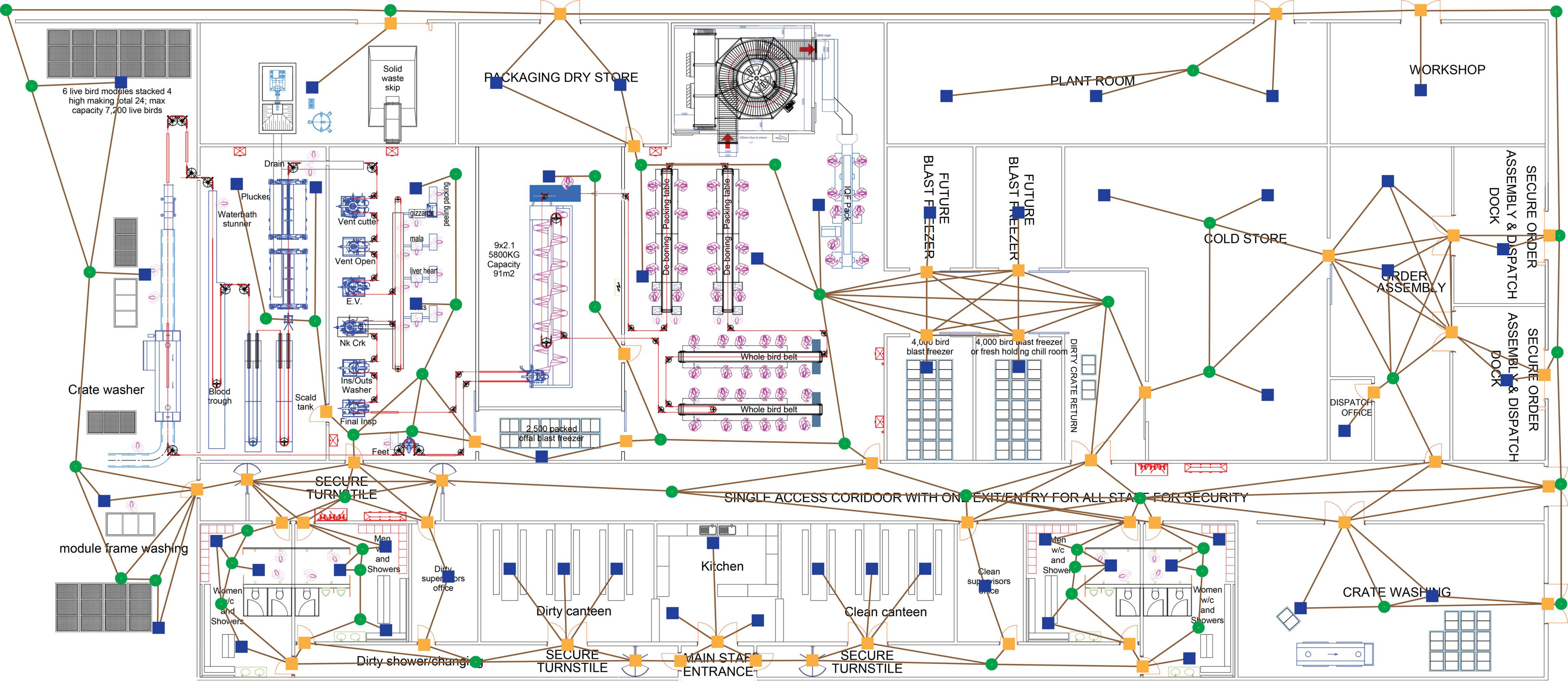} 
  \caption{The \textit{indoor mobility graph} constructed from the factory layout in Fig. \ref{fig:factorySensors}. The dark blue squares are \textit{possible end destinations}. The yellow circles are \textit{doors}. The green squares are \textit{path points}, which define the mobility paths roots can take.}
  \label{fig:factoryMobilityPath}
  \vspace{-1em}
\end{figure*}
\subsubsection{Study of Mobile Networks}
\label{exp_mobility}
The performance of BRPL has been evaluated in networks where mobile devices exist. We managed to obtain a 500mx250m layout for a poultry processing factory from \cite{poultryfactory}, shown in Fig. \ref{fig:factorySensors}. The transmission range is adjusted to 30m to simulate indoor radio limitation. 130 nodes has been semi-randomly placed, \textcolor{black}{generating UDP packets which are required to} be collected by the mobile roots. From the layout, an \textit{indoor mobility graph} has been created as shown in Fig. \ref{fig:factoryMobilityPath}. The graph defines all possible destinations and paths that mobile roots can take while traveling within the factory. The doors are assumed to be always open during the simulations. A custom Cooja plugin has been developed to import the mobility graph to Cooja. In the simulations, each root has 9 preferred destinations (randomly selected during the bootstrap phase). \textcolor{black}{A root travels to one of its preferred destinations 90\% of the time. For} the other 10\%, it randomly chooses a destination from the indoor mobility graph. When a root reaches its destination, it waits for one second, selects a new destination, and the process repeats itself. The travel speed of all roots is set to 1.4 m/s with 0.2 variance \textcolor{black}{to simulate average man walking speed}.

\begin{figure}[h!]
\centering
\subfigure[packet loss.]{
		\includegraphics[trim=60 20 50 10,clip=true,angle=0,width=0.42\textwidth] {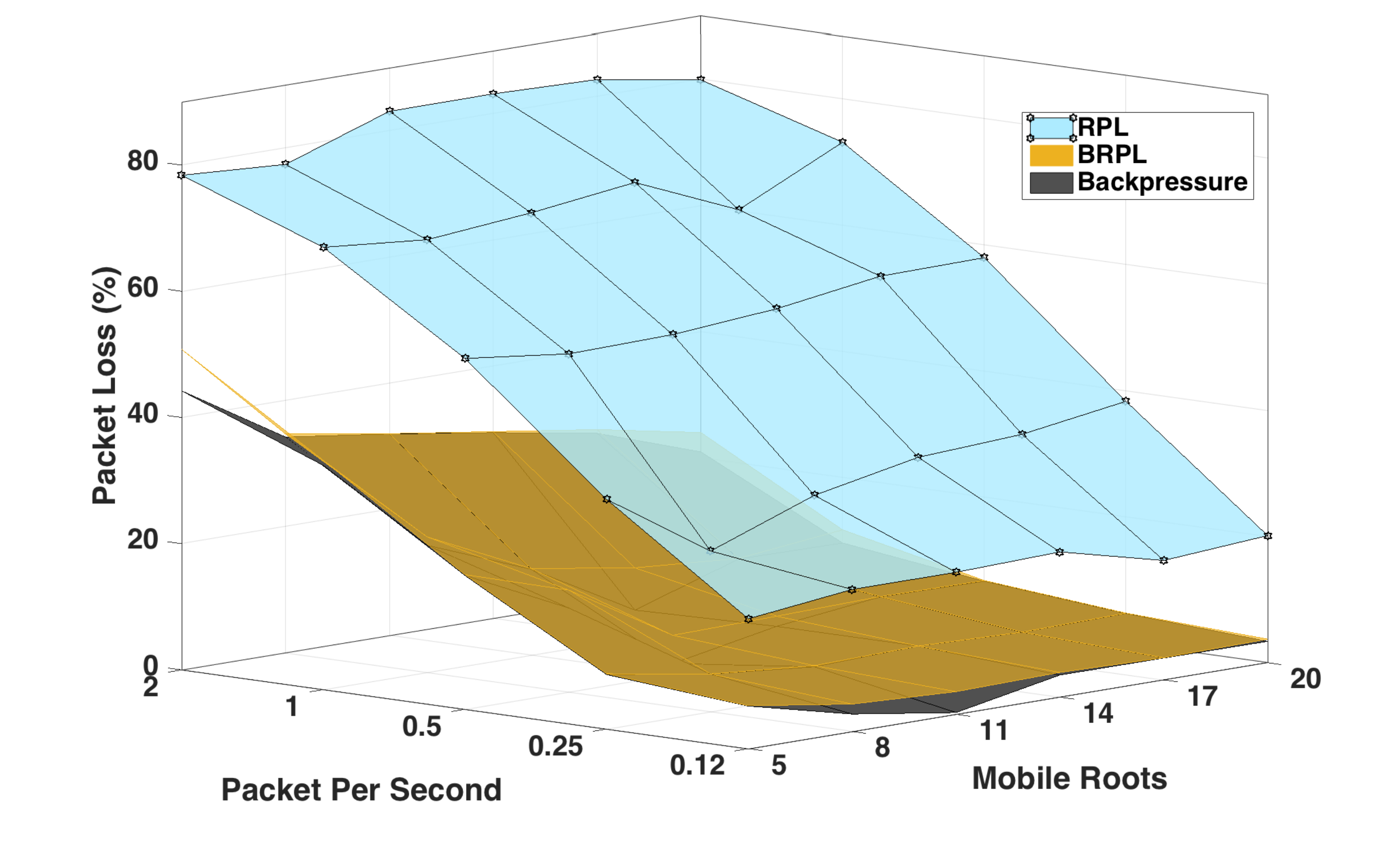} 
		\label{fig:factoryPacketLoss}
	}
\subfigure[end-to-end delay.]{
		\includegraphics[trim=60 20 50 10,clip=true,angle=0,width=0.42\textwidth] {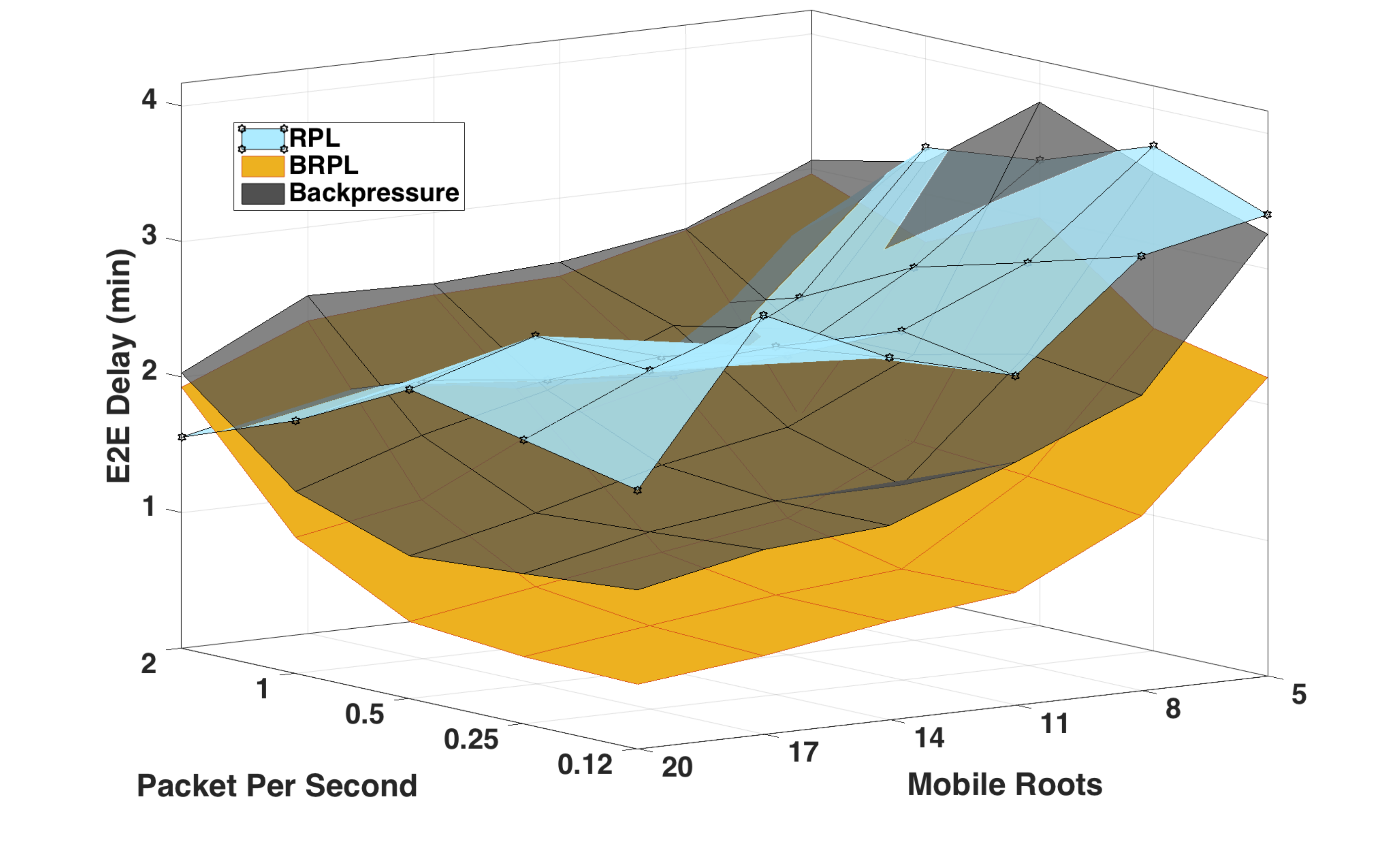} 
		\label{fig:factoryPacketDelay}
	}
\subfigure[communication overhead (in $10^6$ transmissions). ]{
		\includegraphics[trim=60 20 50 10,clip=true,angle=0,width=0.42\textwidth] {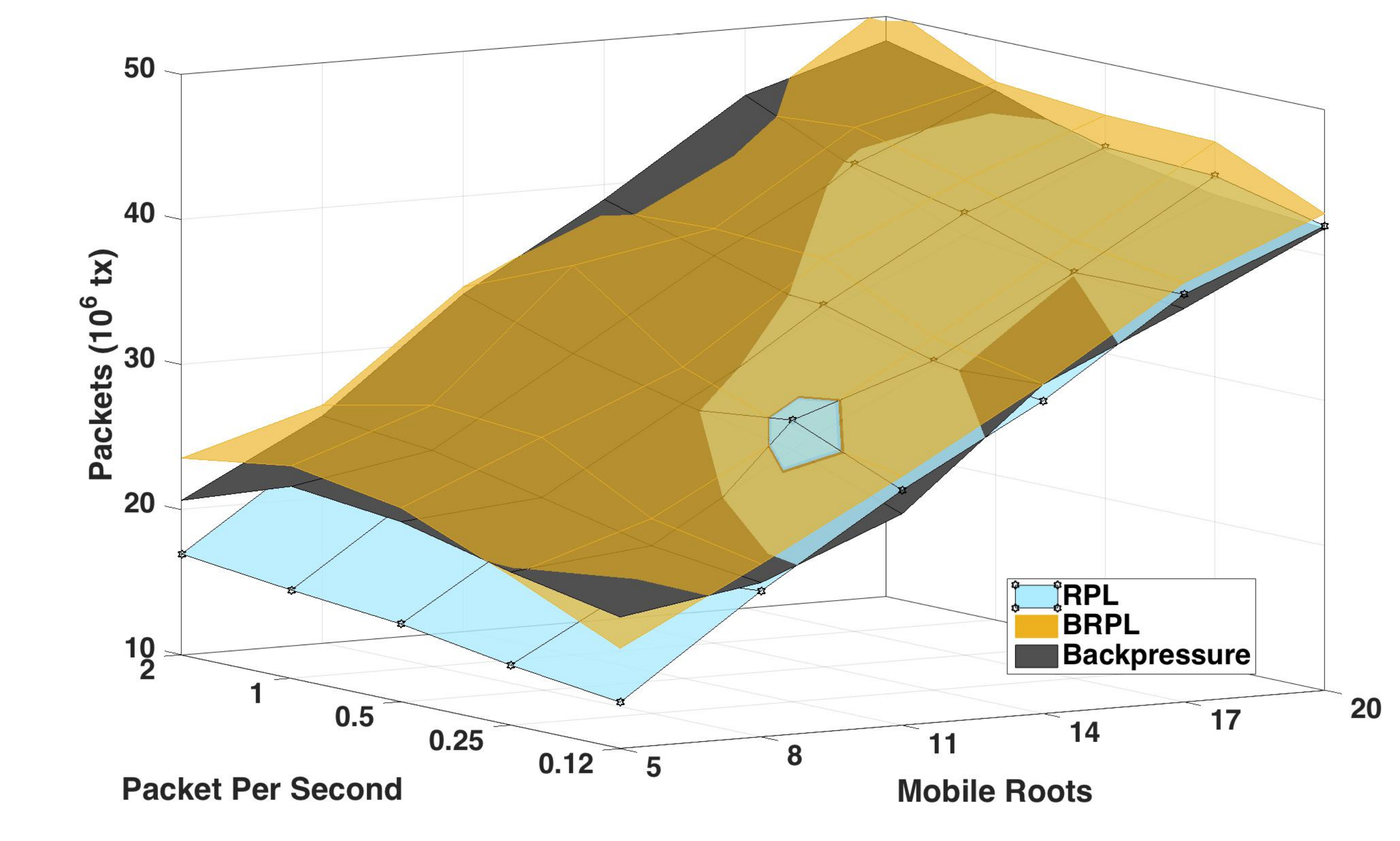} 
		\label{fig:factoryPacketControl}
	}
	\caption{The performance of BRPL, RPL and backpressure routing in a mobile network simulated based on the layouts in Fig. \ref{fig:factorySensors} and \ref{fig:factoryMobilityPath}. BRPL outperforms RPL in terms of packet loss and average end-to-end delay. }
    \vspace{-1em}
\end{figure}

Fig. \ref{fig:factoryPacketLoss} shows the packets loss for BRPL, RPL, and backpressure routing under various network and traffic conditions. As seen, the packet loss of BRPL is between 1.2 and 20 times less than RPL. RPL relies on the ETX OF alone, commonly known to be unsuitable for mobile networks. Outdated DAGs and link breakage were the main reasons for losing 70\% to 80\% of the data packets in RPL. BRPL and backpressure routing have relatively close performance (usually within $\leq 2\%$ differences), and thanks to adaptive routing operations they utilizes mobile resources more effectively than RPL. As observed in Fig. \ref{fig:factoryPacketLoss}, the more mobile roots are available, the less packet loss is expected from the network running BRPL or backpressure routing.

Fig. \ref{fig:factoryPacketDelay} presents the average end-to-end delays for the delivered data packets. RPL has usually worst performance than the other two. The outdated DAGs causes nodes to send packets to a destination where the roots no longer exist and the ETX OF does not cope well with utilizing opportunistic resources. BRPL has better performance than backpressure routing as it tends to, especially under light traffic loads, to utilizing OF reduces routing loops.

The communication overhead for the three routing schemes are similar as seen in Fig. \ref{fig:factoryPacketControl}. The Trickle algorithm and the IPv6 Neighbor Discovery Service causes the mobile roots to continuously transmit DIO and IPv6 ND control messages. These messages are the main reasons for the large communication overhead.

\subsubsection{Study of Adaptive Balancing in QuickTheta}

The purpose of this set of simulations is to show QuickTheta performance at runtime. We want to demonstrate how QuickTheta can effectively find the best possible balance between maximizing throughput and minimizing RPL OF. To do so, the performance of QuickTheta is compared to the drift-plus-penalty routing technique \cite{neely2003dynamic} under various $V$ settings.

In order to easily visualize the dynamics of the $\theta$ parameter, a simple grid topology is used. The topology has 100 nodes with spacing of 30 meters between adjacent nodes. Radio transmission range is set to 50 meters. The network has 98 sensor nodes and two roots (root1 and root2). Root1 is set to be plugged in a power line with unlimited amount of energy. Root2, on the other hand, runs on battery with a finite amount of energy. Root1 and root2 are respectively deployed in the top left corner (grid cell (0,0)) and the bottom right (grid cell (10,10)). The network uses a custom OF such that it is designed to reduce maintenance cost, and therefore always prefer to use root1 for the data traffic. Based on the implementation of the OF, root2 will be used only when root1 is not available.

\begin{figure*}[h!]
\centering
	\subfigure[$\theta$ for a grid topology under two different data rates (PPS=1 and PPS=2).]{
		\includegraphics[trim=5 40 30 10,clip=true,angle=0,width=0.33\textwidth] {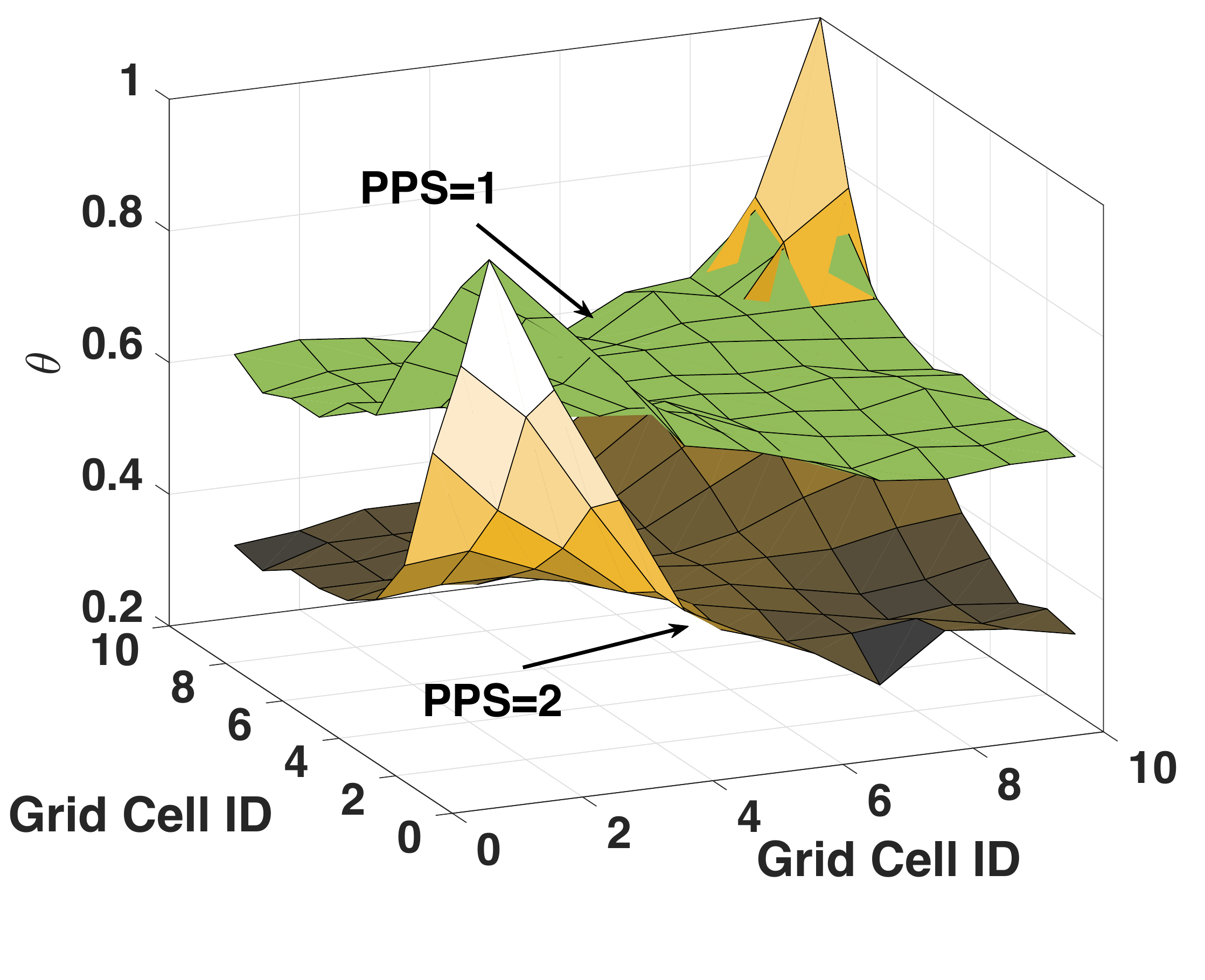} 
		\label{fig:grid_theta}
	}~
	\subfigure[when data rate is 1 packet per second, BRPL has similar performance to the drift-plus-penalty scheme with V=5.]{
		\includegraphics[trim=0 0 0 0,clip=true,angle=0,width=0.33\textwidth] {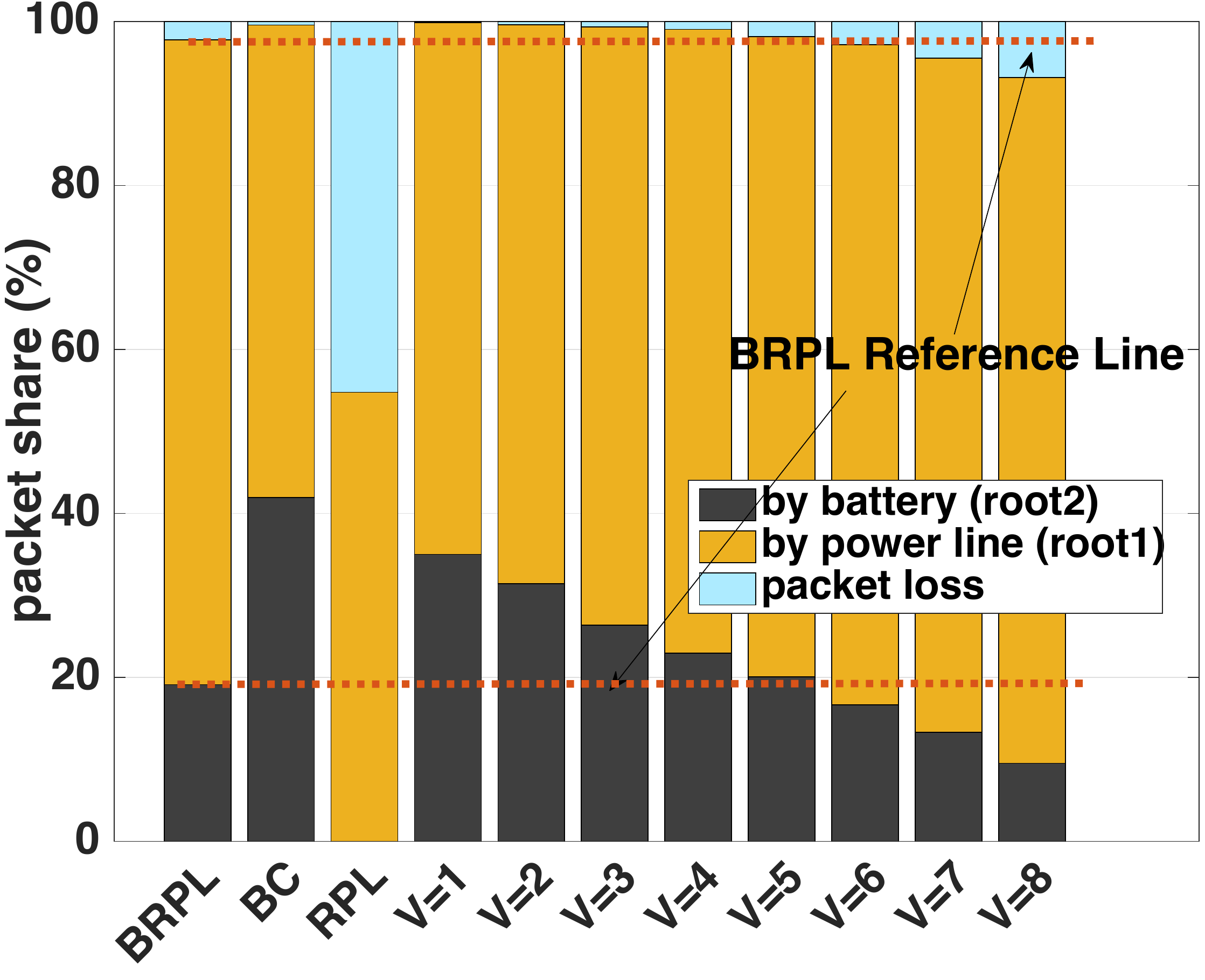} 
		\label{fig:grid_rx1}
	}~
	\subfigure[when data rate is 2 packet per second, BRPL has similar performance to the drift-plus-penalty scheme with V is somewhere in the range (3,4).]{
		\includegraphics[trim=0 0 0 0,clip=true,angle=0,width=0.33\textwidth] {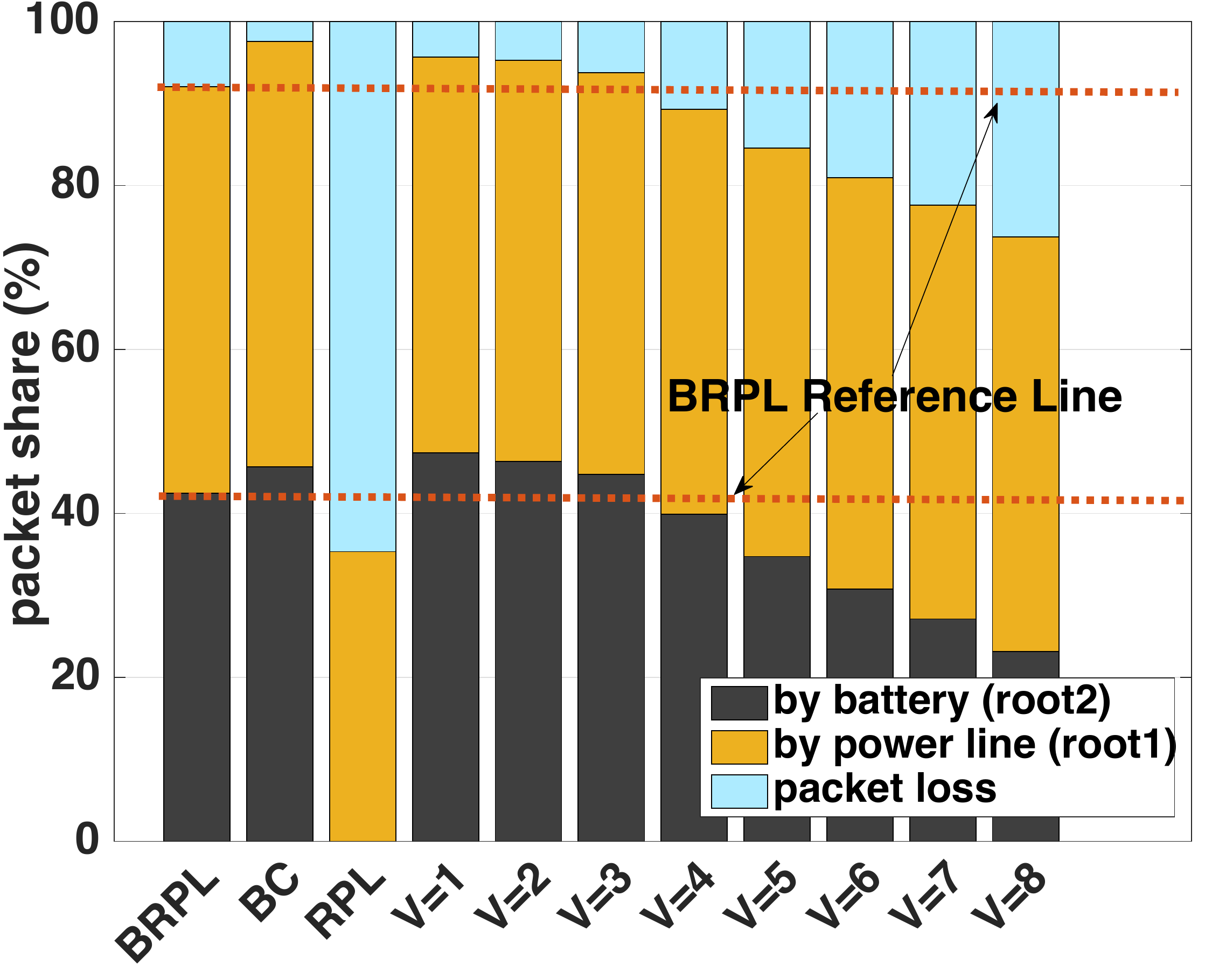} 
		\label{fig:grid_rx2}
	}
	\caption{The performance of for BRPL, backpressure routing, RPL, and drift-plus-penalty scheme with various $V$ settings in a grid topology. The network has two roots deployed at the corners. (a) shows the average $\theta$ values for the nodes under two different data rates. (b) and (c) present the observed packet loss and the amount of data packets forwarded to each root.}
    \vspace{-1em}
\end{figure*}

Fig. \ref{fig:grid_theta} shows the $\theta$ parameter for the network under two different traffic loads. Depending on the traffic congestion levels, the QuickTheta algorithm tends to converge the $\theta$ parameter into a relatively fixed range. For example, in Fig. \ref{fig:grid_theta} when the data rate is 2 packet per second, $\theta$ for the most nodes tends to be in the range [0.25,0.35]. However, the nodes, which are closer to the roots, will have a higher $\theta$ (closer to 1) since the traffic levels are less in these spots than the rest of the network.

The drift-plus-penalty technique proposed in \cite{neely2003dynamic} has been implemented to provide another approach to combine RPL OF with backpressure routing. However as going to be discussed in Section \ref{sec:related_work}, the key challenge in utilizing this technique is that it requires manual tuning for the $V$ parameter, which is the tradeoff parameter between throughput and achieving the optimal solution for the penalty function (i.e. RPL OF in our case). Finding the optimal value for the $V$ parameter is  difficult, especially in multi-user IoT systems. $V$ tightly depends on the expected traffic levels from the application layer. The traffic levels in IoT can be highly dynamic and time-varying. For this set of simulations, we manually tested the network under various $V$ values.

Fig. \ref{fig:grid_rx1} shows the performance BRPL, RPL, backpressure and drift-plus-penalty technique with different $V$ settings. Here the data rate is fixed to 1 packet per second. Backpressure routing provides a costly allocation as 41\% of data packets are forwarded to root2 (i.e., 41\% of the data is gathered using the stored energy from root2's batteries). This is expensive in terms of maintenance costs. RPL, on the other, always tends to forward all the data packets to root1 using OF optimal paths, which results in 45\% packet loss. BRPL first tries to greedily forward data to root1. However as the $\theta$ parameter gradually increases because of network traffic congestion, it utilizes root2 and other suboptimal paths to upload the rest of the packets. The result is root1 gets 78.6\% of the data packets and root2 gets 19\% of the data packets. From Fig. \ref{fig:grid_rx1}, it is clear that BRPL performance is very similar to the drift-plus-penalty technique when $V$ is close to 5. If $V$ is below 5, the allocation is not as optimal since more data packets are forwarded via energy stored in root2's batteries. If $V$ is above 5, a higher packet loss then will be observed. A similar story can be seen when the data rate is 2 PPS. As shown in Fig. \ref{fig:grid_rx2}, BRPL sends 42\% and 49\% of data packets to root1 and root2 respectively. A similar performance can be found when $V$ equalizes to a certain point in the range (3,4). Increasing $V$ beyond that point will causes packet loss as the routing scheme aggressively minimizes the OF. Decreasing $V$ from this range will cause more data forwarding towards root2, which is not optimal either. Therefore, QuickTheta in both cases managed to find the best practically possible tradeoff between throughput and minimizing the OF.

\vspace{-0.6em}
\section{Related Work}
\label{sec:related_work}
\subsection{RPL} The proposed extension in \cite{ancillotti2014reliable} addresses the data reliability problem in RPL. Similarly, ORPL \cite{duquennoy2013} incorporates opportunistic routing with RPL to achieve low latency, robustness, and good scalability. On the other hand, \cite{pavkovic2011multipath} provides a modified MAC layer for supporting multipath forwarding. Kalman positioning and the Corona mechanism have been used in KP-RPL\cite{barceloaddressing} and Co-RPL \cite{gaddour2014co}, respectively, to address the issue of mobility support in RPL. BRPL relies on the smart and smooth switching between RPL and backpressure routing to enhance mobility support. This results in a lightweight solution that does not require any assumption or knowledge about the underline mobility pattern of the network. In addition, this paper presents a modular framework. The implementation of QuickBeta can easily be adjusted to incorporate other mobility metrics, including Kalman-based and Corona-based metrics. Enhancing RPL in terms of mobility, throughput and traffic adaptability all together has not yet been examined by the current literature.

\textcolor{black}{We consider QU-RPL, recently proposed in \cite{kim2016load}, to be the closest to our work. The authors here combine queue information with the OF0 to improve the \textit{load-balancing} of RPL routing in heavy-traffic networks.
BRPL also exploits queue backlogs during routing decision making. However,  BRPL does not rely on the \textit{propagation} of  queue metadata in multi-hop paths. Instead, routing decisions in BRPL are performed based on \textit{pure hop-by-hop} queue information, similar to backpressure routing. Clearly, this makes BRPL a more dynamic routing solution, applicable not only for heavy-traffic networks, but also for networks with highly time-varying traffic loads and topologies, such as networks with node mobility which are sadly ignored in QU-RPL. In addition,  BRPL eliminates the need for having a naive \textit{propagation reduction factor}, $\lambda$ in QU-RPL. BRPL also offers an attractive solution to adaptively tune the sensitivity of queues. Manually setting these parameters as suggested in QU-RPL is  challenging in dynamic, large-scale IoT systems. Furthermore unlike QU-RPL, BRPL is not limited to OF0. In fact, any RPL OF can be used with BRPL since data queues are completely decoupled from OFs while both are still strongly normalized. This is vital to truly improving the performance of MTR deployments.}

\subsection{Backpressure Routing}
The proposed solution in this work is heavily inspired by the elegant \textit{backpressure} routing \cite{tassiulas1992stability}. Introducing the drift-plus-penalty technique \cite{neely2003dynamic} and combining it with \cite{neely2009intelligent, neely2008fairness,neely2008order} contributions in utility optimal networking provided a theoretical framework for backpressure-based stochastic optimization. This framework has been used in a wide range of applications including power control \cite{yang2013distributed,li2010energy}, selfish data relays \cite{yang2013selfish}, sensor networks \cite{moeller2010routing}, and mobile networks \cite{neely2005dynamic,shusen2015}. However, utilizing this framework directly in RPL for dynamic IoT systems is not practical. The framework has parameter tuning issues. The $V$ parameter sets the tradeoff between queue backlogs and \textit{penalty/objective} function optimization. Users are required to set the $V$ parameter based on the expected traffic level from the application layer. This is challenging because the routing layer must have certain assumptions about OFs and expected data traffic, which is technically difficult, or even impossible, in LLNs with event-driven applications. A network may need to serve multiple concurrent  \textcolor{black}{users} with heterogeneous time-varying bandwidth demands. \textcolor{black}{This work presents QuickTheta as a practical solution to this problem.}

Many efforts have been made to make backpressure practical. Various techniques have been proposed to improve the performance of backpressure in terms of the reduction of memory overhead \cite{bui2009novel,athanasopoulou2013back} and delay \cite{ji2013throughput, moeller2010routing}. \textcolor{black}{However, to our knowledge} this is the first work that utilizes backpressure routing in hybrid networks where some nodes use backpressure routing while others do not.

\vspace{-0.5em}
\section{Conclusion}
\label{sec:conclusion}

This work addresses three key limitations of RPL: low throughput, poor adaptability to time-varying data traffic loads and lack of support for node mobility.
To this end, we develop BRPL, a backward-compatible extension for RPL, which can adaptively and smoothly switch between RPL and backpressure routing depending on network conditions.
We present \textit{QuickBeta} and \textit{QuickTheta}, two adaptive online algorithms for BRPL, to respectively support node mobility and  balance the tradeoff between network throughput and RPL OF minimization. Through extensive experiments driven by real-world testbed and cloud-based simulations, we show that BRPL works seamlessly with RPL and achieves significant performance improvements in terms of network throughput with 60\% packet loss reduction at a minimum in mobile networks.
An interesting future direction is to study resource fairness issues among multiple coexisting DAGs in RPL and BRPL.
\vspace{-0.5em}
\section*{ACKNOWLEDGMENT} 
This work is sponsored by the Ministry of Higher Education and Scientific Research in Kurdistan/Iraq, Intel Corporation, China `1000 Young Talents Program', and `Young Talent Support Plan' of Xi'an Jiaotong University. We thank the associate editor and anonymous reviewers for their helpful and insightful comments and suggestions, which significantly contribute to improving the paper.
\vspace{-1.0em}
\bibliographystyle{IEEEtran}
\bibliography{IEEEfull, main.bbl}

\vspace{-3.5em}
\begin{IEEEbiography}[{\includegraphics[width=25mm,height=32mm,clip,keepaspectratio]{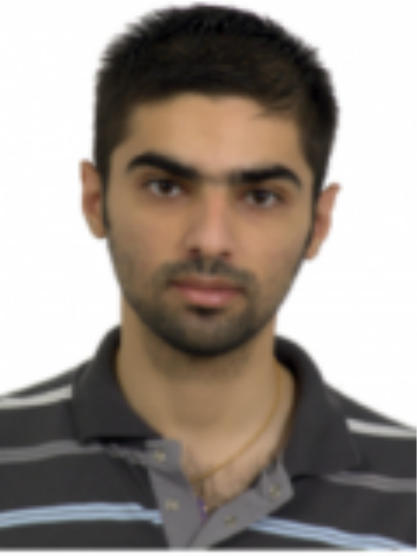}}]{Yad Tahir}
received the B.Sc. degree with Honors in Computer Science from University of Sulaimani, Iraq, in 2008, and the M.Sc in Software Engineering with Distinction from Heriot-Watt University, UK, in 2010.  He has recently finished his Ph.D in Computing at Imperial College London, UK. His research interests include resource management, network control and optimizations, Internet of things, software and system engineering.
\end{IEEEbiography}
\vspace{-3em}
\begin{IEEEbiography}[{\includegraphics[width=25mm,height=32mm,clip,keepaspectratio]{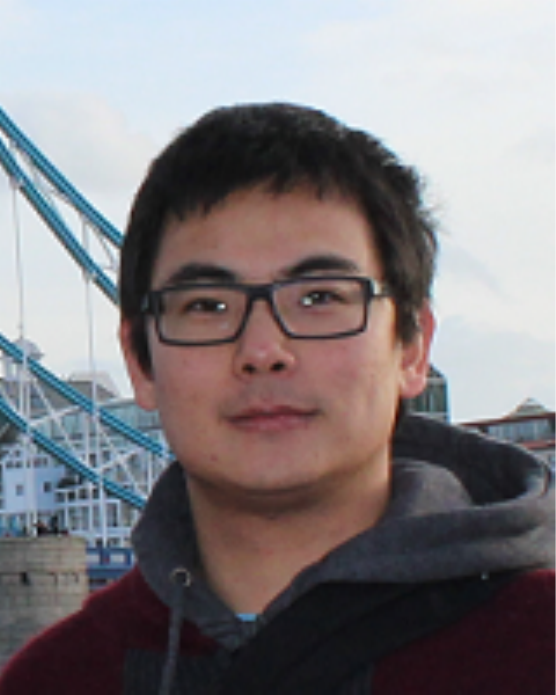}}]{Shusen Yang} received his PhD in Computing from Imperial College London in 2014. He is currently a professor in the Institute of Information and System Science at Xi'an Jiaotong University (XJTU). Before joining XJTU, Shusen worked as a Lecturer (Assistant Professor) at University of Liverpool from 2015 to 2016, and a Research Associate at Intel Collaborative Research Institute ICRI from 2013 to 2014. His research interests include mobile networks, networks with human in the loop, and data-driven networked systems. Shusen achieves  "1000 Young Talents Program" award, and holds an honorary research fellow at Imperial College London.  Shusen is a senior member of IEEE and a member of ACM.
\end{IEEEbiography}
\vspace{-3em}
\begin{IEEEbiography}[{\includegraphics[width=25mm,height=32mm,clip,keepaspectratio]{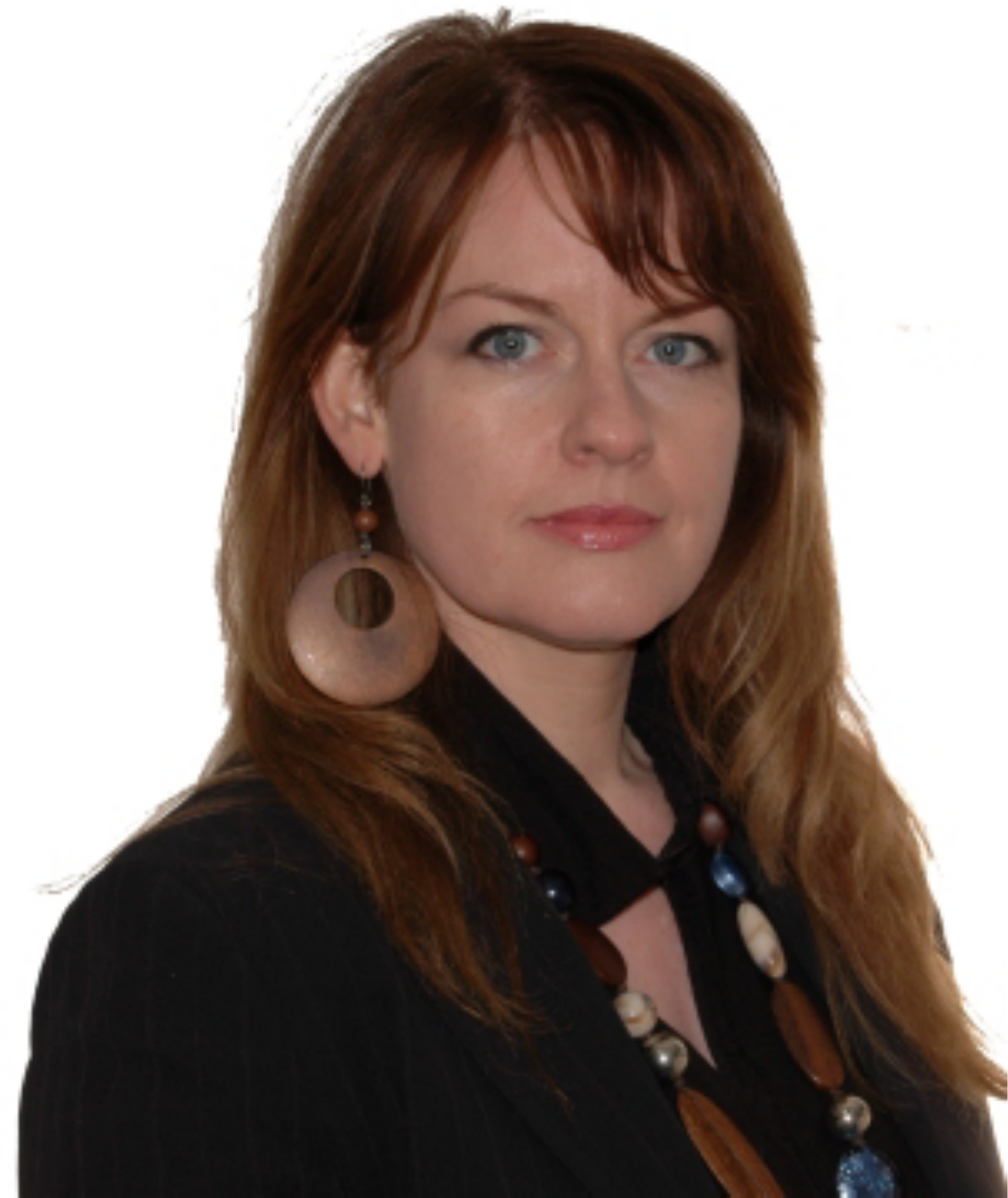}}]{Julie A. McCann }
is a Professor in Computer Systems at Imperial College. Her research centers on highly decentralized and scalable algorithms for spatial computing systems e.g. sensor networks. She leads both the Adaptive Embedded Systems Engineering Research Group and the Intel Collaborative Research Institute for Sustainable Cities, and is  working with NEC and others on substantive smart city projects. She has received significant funding though bodies such as the UK's EPSRC, TSB and NERC as well as various international funds, and is an elected peer for the EPSRC. She has actively served on, and chaired, many conference committees and is currently Associative Editor for the ACM Transactions on Autonomous and Adaptive Systems. She is a Fellow of the BCS.
\end{IEEEbiography}

\end{document}